\documentclass[sigconf,nonacm]{acmart}
\renewcommand\footnotetextcopyrightpermission[1]{}
\AtBeginDocument{%
  \providecommand\BibTeX{{%
    Bib\TeX}}}

\setcopyright{acmlicensed}
\copyrightyear{2026}
\acmYear{2026}
\acmDOI{XXXXXXX.XXXXXXX}
\acmConference[SIGIR '26]{ACM SIGIR Conference 2026}{July 20--24,
  2026}{Melbourne | Naarm, Australia}
\acmISBN{978-1-4503-XXXX-X/2018/06}

\usepackage[linesnumbered,ruled,vlined]{algorithm2e}
\usepackage{multirow}
\usepackage{booktabs}
\usepackage{threeparttable}
\usepackage{amsmath}
\usepackage{amsthm}
\newtheorem{theorem}{Theorem}
\newtheorem{proposition}{Proposition}
\newtheorem{definition}{Definition}

\usepackage{graphicx}
\usepackage{subcaption}

\begin{document}

\title{Cohesive Group Discovery in Interaction Graphs under Explicit Density Constraints}


\author{Yu Zhang}
\affiliation{%
  \institution{Peking University}
  \state{Beijing}
  \country{China}}
\email{yuzhang.cs@stu.pku.edu.cn}

\author{Yilong Luo}
\author{Mingyuan Ma}
\affiliation{%
  \institution{Peking University}
  \state{Beijing}
  \country{China}}

\author{Yao Chen}
\affiliation{%
  \institution{Beijing Capital International Airport Co., Ltd.}
  \state{Beijing}
  \country{China}}

\author{Enqiang Zhu}
\affiliation{%
  \institution{Guangzhou University}
  \city{Guangzhou}
  \country{China}}

\author{Jin Xu}
\affiliation{%
  \institution{Peking University}
  \state{Beijing}
  \country{China}}

\author{Chanjuan Liu}
\affiliation{%
  \institution{Dalian University of Technology}
  \state{Dalian}
  \country{China}}

\renewcommand{\shortauthors}{Zhang et al.}

\begin{abstract}
Discovering cohesive groups is a fundamental primitive in graph-based recommender systems, underpinning tasks such as social recommendation, bundle discovery, and community-aware modeling. In interaction graphs, cohesion is often modeled as the $\gamma$-quasi-clique, an induced subgraph whose internal edge density meets a user-defined threshold $\gamma$. This formulation provides explicit control over within-group connectivity while accommodating the sparsity inherent in real-world data. This paper presents EDQC, an effective framework for cohesive group discovery under explicit density constraints. EDQC leverages a lightweight energy diffusion process to rank vertices for localizing promising candidate regions. Guided by this ranking, the framework extracts and refines a candidate subgraph to ensure the output strictly satisfies the target density requirement. Extensive experiments on 75 real-world graphs across varying density thresholds demonstrate that EDQC identifies the largest mean $\gamma$-quasi-cliques in the vast majority of cases, achieving lower variance than the state-of-the-art methods while maintaining competitive runtime. Furthermore, statistical analysis confirms that EDQC significantly outperforms the baselines, underscoring its robustness and practical utility for cohesive group discovery in graph-based recommender systems.
\end{abstract}





\maketitle

\section{Introduction}\label{introduction}

Cohesive group discovery plays a fundamental role in graph-based recommender systems, enabling tasks such as social recommendation~\cite{alina2014social}, bundle discovery~\cite{nguyen2024bundle}, and community-aware modeling~\cite{guan2021community}. 
In many recommender settings, user--item interactions or user--user relations are naturally modeled as interaction graphs, where vertices denote users or items and edges capture affinity signals such as co-consumption or behavioral similarity~\cite{schafer2001commerce,zhou2025robust}. 
This representation encompasses various network types, including social, communication, citation, collaboration, and web graphs.
Within such graphs, cohesive groups are often instantiated as dense induced subgraphs, supporting candidate expansion, group-aware ranking, and coordinated abnormal behavior detection.

Among dense subgraph models, the \emph{clique} is the most stringent, requiring every pair of vertices to be connected~\cite{saito2007large}. 
However, in real-world networks, noise, missing edges, and incomplete connectivity make exact cliques rare. 
To address this, several relaxed models have been proposed, including \emph{quasi-cliques}~\cite{abello2002massive,pei2005mining}, \emph{$k$-clubs}~\cite{mokken1979cliques}, and \emph{$k$-plexes}~\cite{seidman1978graph}. 
In particular, the edge-density-based formulation, often referred to as the \emph{$\gamma$-quasi-clique}, is widely adopted for its intuitive and explicit notion of structural cohesion~\cite{chen2021nuqclq,pang2024similarity,liu2024optimization}.

Given an undirected graph $G$ and an edge-density threshold $\gamma \in (0,1]$, the Maximum Quasi-Clique Problem (MQCP) asks for a largest vertex set $S$ such that the induced subgraph $G[S]$ has edge density at least $\gamma$. 
Throughout the paper, \emph{density} refers to \emph{edge density}, and \emph{quasi-clique} refers to the $\gamma$-quasi-clique unless otherwise stated.
Unlike degree-based criteria used in FastQC~\cite{konar2024optimal} and IterQC~\cite{liu2008effective}, MQCP enforces an explicit, user-specified density threshold. 
It is also distinct from community detection, which typically optimizes a global objective (e.g., modularity or conductance) to partition or cover the entire graph. 
MQCP instead extracts a single dense local vertex set, aligning with settings where one seeks a single cohesive set of users or items rather than a full partition.
In recommender applications, $\gamma$ serves as a minimum cohesion requirement that explicitly controls within-group connectivity. 
Hence, methods that do not enforce a user-specified $\gamma$ may produce subgraphs that fail to satisfy the target density, limiting their utility as reliable constraint-aware components for recommendation tasks.

MQCP is NP-hard even for a fixed $\gamma$~\cite{pattillo2013maximum} and lacks the \emph{hereditary property} used for pruning in clique-finding algorithms~\cite{xia2025maximum}. In particular, a subset of a $\gamma$-quasi-clique is not necessarily a $\gamma$-quasi-clique. Despite these challenges, MQCP remains practically relevant, since $\gamma$-quasi-cliques capture compact interest groups and suspicious behavioral clusters in real-world graphs with dense but imperfect connectivity.

\begin{figure}[tb]
\centering
\includegraphics[width=0.55\columnwidth]{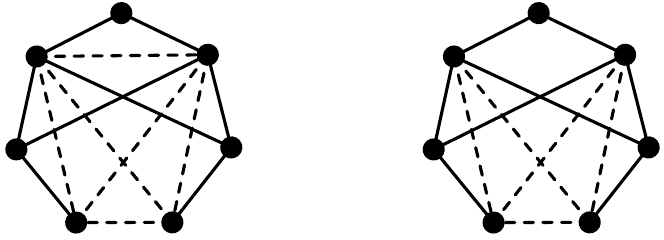} 
\caption{A size-4 clique and a 0.8-quasi-clique. Dashed edges indicate observed interaction affinities.}
\label{fig:intro}
\end{figure}

Figure~\ref{fig:intro} illustrates a typical situation in interaction graphs. In the left subgraph, the dashed edges form a size-4 clique, whereas in the right subgraph one edge is missing but the dashed edges still constitute a $0.8$-quasi-clique. In practice, interactions are often noisy or partially observed, so nearly-complete groups like this should still be treated as coherent interest groups. This motivates studying $\gamma$-quasi-cliques as an explicit density-constrained model of structural cohesion.

Despite recent progress, discovering quasi-cliques in large graphs remains challenging due to both scale and combinatorial complexity. 
Exact MQCP solvers based on integer programming and branch-and-bound~\cite{uno2010efficient,pastukhov2018maximum,ribeiro2019exact,rahman2024fast} provide optimality guarantees but are restricted to small graphs in practice. Consequently, heuristics are often preferred for real-world graphs, and existing methods can be roughly grouped into four categories.

First, \emph{greedy expansion} methods grow candidates with lightweight refinement~\cite{tsourakakis2013denser,oliveira2013construction}. While fast, they are prone to getting trapped in poor local optima. 
Second, \emph{metaheuristic} frameworks such as TSQC and OBMA~\cite{djeddi2019extension,zhou2020opposition} enhance exploration using strategies like tabu search or opposition-based learning, but typically incur higher computational overhead, which limits scalability on large graphs. 
Third, \emph{local search} algorithms, notably NuQClq and its variants~\cite{chen2021nuqclq,liu2024optimization}, iteratively refine candidates with adaptive move rules (e.g., bounded configuration checking); however, their performance can vary substantially across random seeds on large-scale instances. This seed sensitivity hinders reproducibility and yields unstable outputs across runs, which is undesirable in recommender systems.
Fourth, \emph{similarity-based} heuristics such as NBSim and FastNBSim~\cite{pang2024similarity} exploit neighborhood overlap for efficiency, but they do not take a user-specified edge-density threshold $\gamma$ as input. Instead, density is controlled indirectly via similarity parameters, so the output is not guaranteed to satisfy a target $\gamma$. As a result, their outputs may be dense under an internal similarity criterion yet still violate a required edge-density threshold, making them unsuitable when $\gamma$ is a hard constraint.
Among these categories, local search and similarity-based heuristics are often among the strongest baselines for mining large quasi-cliques. In particular, representative methods include NBSim/FastNBSim~\cite{pang2024similarity} and the local search algorithms NuQClqF1/NuQClqR1~\cite{liu2024optimization}. 
However, these methods either exhibit sensitivity to random seeds, or do not take a user-specified density threshold $\gamma$ as an explicit constraint and thus cannot guarantee meeting a target $\gamma$.

Energy diffusion provides a natural mechanism to improve the stability of quasi-clique discovery. Diffusion-based techniques such as PageRank~\cite{page1999pagerank} and the heat kernel~\cite{chung2007heat} have been effective for structure-aware tasks, including vertex ranking and local community detection~\cite{kloster2014heat,gleich2015multilinear}. A simple intuition is that short-range propagation tends to retain more mass within well-connected regions due to repeated exchanges along internal edges, which can be used to produce a locality-aware ranking from a source vertex. However, classical diffusion models are designed to approximate a probability distribution and are not tailored to extracting a vertex set that explicitly satisfies a density threshold $\gamma$.

To bridge this gap, we present EDQC, an effective framework that leverages energy diffusion for quasi-clique discovery. By utilizing lightweight primitives and sidestepping the iterative overhead often associated with metaheuristic search, EDQC is designed to be practical across different graph sizes.
EDQC first runs an adaptive energy diffusion process to obtain an energy ranking, where the release rate of each vertex is adjusted by the energy in its neighborhood to concentrate scores within locally cohesive regions. 
While the ranking highlights promising vertices, it does not define a subgraph boundary. We therefore apply a conductance-based sweep cut to extract a well-separated candidate region, followed by a refinement step that prunes low-degree vertices and greedily adds qualifying vertices. The refinement enforces feasibility, ensuring that the output meets the user-specified density threshold $\gamma$.

Our main contributions are summarized as follows:
\begin{itemize}
    \item To our knowledge, we are the first to integrate energy diffusion into a heuristic framework for MQCP, using adaptive diffusion-based scoring to localize cohesive regions in interaction graphs at scale.
    \item We design a diffusion-guided framework that converts local rankings into feasible $\gamma$-quasi-cliques through conductance-based extraction and density-constrained refinement, ensuring that the output satisfies the user-specified density threshold $\gamma$.
    \item We evaluate EDQC on 75 real-world graphs with ablation studies and statistical significance tests, showing that it finds larger quasi-cliques than strong baselines on most datasets, with lower variance across runs and competitive runtime.
\end{itemize}

\section{Preliminaries}
All graphs are undirected and simple (no loops or parallel edges).
Let $G=(V,E)$ denote a graph with vertex set $V$ and edge set $E$, where $n=|V|$ and $m=|E|$.
For a vertex $u\in V$, let $N(u)$ be its neighborhood, $N[u]=N(u)\cup\{u\}$ its closed neighborhood, and $d(u)=|N(u)|$ its degree.

For a set $S\subseteq V$, its volume is $\mathrm{vol}(S)=\sum_{u\in S} d(u)$.
Let $G[S]$ be the induced subgraph with vertex set $S$ and edge set
$E(S)=\{\{u,v\}\in E \mid u,v\in S\}$.
Let $m(S)=|E(S)|$ be the number of internal edges in $G[S]$, and for $u\in S$ define the internal degree $d_S(u)=|N(u)\cap S|$.
The edge density of $G[S]$ is
\[
\eta(G[S])=
\begin{cases}
\dfrac{2m(S)}{|S|(|S|-1)}, & |S|\ge 2,\\
0, & \text{otherwise}.
\end{cases}
\]

\begin{definition}[$\gamma$-Quasi-Clique]
Given $G=(V,E)$ and a density threshold $\gamma\in(0,1]$, a vertex set $S\subseteq V$ \emph{induces a $\gamma$-quasi-clique} if the induced subgraph $G[S]$ has edge density at least $\gamma$, i.e., $\eta(G[S]) \;\ge\; \gamma.$
For brevity, we also refer to $G[S]$ itself as a $\gamma$-quasi-clique.
\end{definition}

\begin{definition}[Maximum Quasi-Clique Problem]
Given $G=(V,E)$ and a density threshold $\gamma\in(0,1]$, the \emph{Maximum Quasi-Clique Problem (MQCP)} seeks
\[
S^* \in \arg\max_{S\subseteq V}\ \Big\{|S| \ \Big|\ \eta(G[S])\ge \gamma\Big\}.
\]
The induced subgraph $G[S^*]$ is a maximum $\gamma$-quasi-clique. 
When $\gamma=1$, MQCP reduces to the classical Maximum Clique Problem, and for any fixed $\gamma$ the problem is NP-hard and may admit multiple optimal solutions~\cite{pattillo2013maximum}.

\end{definition}

\begin{definition}[Conductance]
Conductance measures how well a vertex set is separated from the rest of the graph and is widely used in local clustering~\cite{andersen2006local}. 
For a non-empty, proper subset of vertices $S \subset V$, the \emph{cut} between $S$ and its complement is defined as $\mathrm{cut}(S, V \setminus S) = |\{\{u,v\} \in E \mid u \in S, v \in V \setminus S\}|$.
The conductance of $S$ is then:
\[
\Phi(S) = \frac{\mathrm{cut}(S, V \setminus S)}{\min(\mathrm{vol}(S), \mathrm{vol}(V \setminus S))}.
\]
A smaller $\Phi(S)$ indicates better separation.
\end{definition}

\section{The EDQC Framework} \label{sec:method}

This section presents EDQC for MQCP under an explicit density threshold $\gamma$.
We use EDQC to denote the proposed framework. Algorithm~\ref{alg:edqc} provides an instantiation of this framework (referred to as the EDQC algorithm when discussing pseudocode and complexity).

EDQC iterates over vertices as diffusion sources in non-increasing order of degree and maintains the best feasible solution found across all processed sources (lines 2--6).
This ordering follows a standard efficiency heuristic~\cite{rossi2014fast} that prioritizes high-degree sources, which tend to yield larger candidate regions early and thus increase the chance of finding a large feasible quasi-clique in fewer iterations.
For each source vertex $v$, EDQC first computes an energy score $f:V\to\mathbb{R}_{\ge 0}$ via an adaptive diffusion procedure, providing a locality-aware ranking of vertices around $v$ (line 3).
It then converts this ranking into a feasible quasi-clique by extracting a candidate set and performing density-constrained refinement to ensure the output satisfies the target threshold $\gamma$ (line 4).
After computing $S$ for the current source, EDQC keeps it if it improves the best solution so far and returns the best $S^*$ at the end (lines 5--7).

\begin{algorithm}[htbp]
    \small
    \caption{EDQC$(G, \gamma, T, \theta)$}
    \label{alg:edqc}
    \KwIn{Graph $G=(V,E)$, density threshold $\gamma$, diffusion rounds $T$, activation threshold $\theta$}
    \KwOut{Vertex set $S^*$ of a large $\gamma$-quasi-clique}
    $S^* \gets \emptyset$\;
    \ForEach{$v \in V$ in non-increasing order of degree}{
        $f \gets$ \textsc{EnergyDiffusion}$(G, v, T, \theta)$\;
        $S \gets$ \textsc{ExtractQuasiClique}$(G, f, \gamma, \theta)$\;
        \If{$|S| > |S^*|$}{
            $S^* \gets S$\;
        }
    }
    \Return{$S^*$}
\end{algorithm}

\subsection{Adaptive Energy Diffusion}
The procedure (Algorithm~\ref{alg:energy-diffusion}) diffuses a unit amount of energy from a source vertex $v$ for $T$ synchronous rounds.
It outputs an energy function $f:V\to\mathbb{R}_{\ge 0}$ that induces a locality-aware ranking around $v$.
The ranking is used for candidate localization, while the density constraint $\gamma$ is enforced later by extraction and refinement.

\begin{algorithm}[t]
\small
    \caption{\textsc{EnergyDiffusion}$(G, v, T, \theta)$}
    \label{alg:energy-diffusion}
    \KwIn{Graph $G=(V,E)$, source vertex $v$, diffusion rounds $T$, activation threshold $\theta$}
    \KwOut{Energy function $f:V\to\mathbb{R}_{\ge0}$}
    Initialize $f(v) \gets 1$ and $f(x) \gets 0$ for $x \neq v$\;
    \For{$t = 1$ \KwTo $T$}{
        $A \gets \{x \in V \mid f(x) > \theta\}$\;
        $f' \gets f$\;
        \ForEach{$u \in A$}{
            \If{$d(u) > 0$}{
                $\alpha(u) \gets \frac{\sum_{w \in N(u)} f(w) + f(u)}{\sum_{w \in N(u)} f(w) + 2f(u)}$\;
                $\delta(u) \gets \alpha(u) f(u)$\;
                Sample $\xi_{uw} \sim \mathrm{Exp}(1)$ for each $w \in N(u)$\;
                $\phi_{uw} \gets \frac{\xi_{uw}}{\sum_{x \in N(u)} \xi_{ux}}$ for each $w \in N(u)$\;
                $f'(u) \gets f'(u) - \delta(u)$\;
                \ForEach{$w \in N(u)$}{
                    $f'(w) \gets f'(w) + \delta(u) \phi_{uw}$\;
                }
            }
        }
        $f \gets f'$\;
    }
    \Return{$f$}
\end{algorithm}

\paragraph{Design rationale.}
EDQC runs a fixed-step local diffusion for $T$ rounds and uses two simple mechanisms.
A vertex is \emph{active} if its current energy exceeds $\theta$, and the active set is $A=\{x\in V\mid f(x)>\theta\}$ (line~3).
First, the released amount from an active vertex is adaptive: for $u\in A$, the algorithm releases $\delta(u)=\alpha(u)f(u)$, where $\alpha(u)$ increases with the total energy in the neighborhood of $u$ (line~7--8).
This adaptivity encourages faster energy exchange when the diffusion remains inside a well-connected region around the source and slows down propagation when the mass reaches low-energy surroundings, which is consistent with diffusion-based local clustering using short-range propagation~\cite{andersen2006local,kloster2014heat,chung2007heat}.
Second, in each round, EDQC updates only active vertices (line~5), which keeps the computation local and avoids spending time on vertices with negligible energy.

Compared to classical diffusion models such as PageRank~\cite{page1999pagerank} and the heat kernel~\cite{chung2007heat}, EDQC does not aim to approximate a stationary distribution.
Instead, it runs for a fixed $T$ rounds and uses the scores to induce a local ranking around the source.

\paragraph{Procedure.}
Algorithm~\ref{alg:energy-diffusion} initializes $f(v)=1$ and $f(x)=0$ for $x\neq v$ (line~1).
In each round, it forms the active set $A$ (line~3) and uses a temporary copy $f'$ to ensure synchronicity across updates (line~4).
For each active vertex $u$ with $d(u)>0$, it computes the adaptive fraction $\alpha(u)$ and the released amount $\delta(u)=\alpha(u)f(u)$ (lines~5--8).
It then stochastically allocates $\delta(u)$ to neighbors by sampling i.i.d.\ exponential variables and normalizing them into weights $\phi_{uw}$ (line~9--10), and updates $f'$ for $u$ and its neighbors accordingly (line~11--13).
Vertices with $d(u)=0$ have no neighbors to receive energy and are therefore skipped, leaving their energy unchanged.
Each round ends by setting $f\gets f'$ (line~14), and after $T$ rounds the algorithm returns the final energy function $f$ (line~15).

\paragraph{Properties.}
We summarize several basic properties of the diffusion procedure.

\begin{proposition}[Energy Conservation]
The total energy is conserved in each round: $\sum_{x \in V} f'(x) = \sum_{x \in V} f(x)$.
\end{proposition}

\begin{proof}
Consider a fixed round and an arbitrary active vertex $u$ where $d(u) > 0$. The synchronous updates associated with $u$ result in a change of $-\delta(u)$ for $u$ and an addition of $\delta(u) \phi_{uw}$ for each neighbor  $w \in N(u)$, satisfying $\sum_{w\in N(u)} \phi_{uw} = 1$. Therefore, $\delta(u) = \sum_{w\in N(u)} \delta(u) \phi_{uw}$. Summing over all active vertices leads to the conclusion that 
$\sum_{x\in V} f'(x) = \sum_{x\in V} f(x)$.
Vertices that are not updated in this round (including those with $f(u)\le \theta$ or $d(u)=0$) remain unchanged and thus do not affect the sum.
\end{proof}

The conservation of energy is a crucial property for the algorithm's stability. It guarantees that the process only redistributes the initial energy, rather than arbitrarily creating or losing it.

\begin{proposition}[Properties of $\alpha(u)$] \label{prop:alpha-bounds}
For an active vertex $u$ with $f(u)>0$ and $d(u)>0$, the adaptive fraction $\alpha(u) \in [1/2, 1)$ and is monotonically increasing with the total energy of its neighbors, $S(u) = \sum_{w \in N(u)} f(w)$.
\end{proposition}
\begin{proof}
We begin by rewriting the formula for $\alpha(u)$ (line~7, Algorithm~\ref{alg:energy-diffusion}) as $\alpha(u) = 1 - \frac{f(u)}{S(u) + 2f(u)}$. Since $S(u) \ge 0$ and $f(u) > 0$ for an active vertex, the denominator $S(u) + 2f(u)$ is strictly positive. The fractional term is bounded by $0 < \frac{f(u)}{S(u) + 2f(u)} \le \frac{f(u)}{2f(u)} = 1/2$. Consequently, $\alpha(u)$ is bounded by $1 - 1/2 \le \alpha(u) < 1 - 0$, which gives $\alpha(u) \in [1/2, 1)$.
For monotonicity, the partial derivative with respect to $S(u)$ is:
\[
\frac{\partial \alpha(u)}{\partial S(u)} = \frac{f(u)}{(S(u)+2f(u))^2} > 0.
\]
This confirms that $\alpha(u)$ is strictly increasing with $S(u)$.
\end{proof}

The adaptive nature of the diffusion is formalized by this proposition. The bounds on $\alpha(u)$ imply that each active vertex always releases at least half, but strictly less than all, of its current energy. Thus, the process continues to make progress in every round without becoming unstable. Moreover, the monotonic dependence of $\alpha(u)$ on the neighbors' total energy means that vertices in high-energy neighborhoods release a larger fraction of energy, so diffusion proceeds more rapidly within dense regions and more slowly outside these dense regions.

\begin{proposition}[Unbiased Distribution in Expectation]
The expected fraction of energy that an active vertex $u$ transfers to a specific neighbor $w \in N(u)$ is uniform across all neighbors, that is, $\mathbb{E}[\phi_{uw}] = 1/d(u)$.
\end{proposition}
\begin{proof}
The weights $\phi_{uw}$ are derived by normalizing i.i.d. exponential random variables $(\xi_{uw})_{w \in N(u)}$. This construction is a standard method for generating a random point from a uniform distribution on the standard simplex. By symmetry, the expectation of each component is identical. Since $\sum_{w \in N(u)} \phi_{uw} = 1$, by linearity of expectation, $\sum_{w \in N(u)} \mathbb{E}[\phi_{uw}] = 1$. As there are $d(u)$ identical terms in the sum, it follows that $d(u) \cdot \mathbb{E}[\phi_{uw}] = 1$, which proves $\mathbb{E}[\phi_{uw}] = 1/d(u)$.
\end{proof}

Conditioned on the current energy $f$, for an active vertex $u$ and any neighbor $w\in N(u)$, the expected update satisfies $\mathbb{E}[f'(w)] = f(w) + \delta(u)/d(u)$. 
Thus, the diffusion is uniform over neighbors in expectation; any energy concentration arises from repeated exchanges along the graph structure (as in short-range diffusion for local clustering~\cite{andersen2006local,kloster2014heat,chung2007heat}).

\begin{theorem}[Complexity of Algorithm~\ref{alg:energy-diffusion}] 
\label{thm:diffusion_complexity}
\textsc{EnergyDiffusion} runs in $O(T(n+m))$ time and uses $O(n+m)$ space.
\end{theorem}
\begin{proof}
The algorithm executes $T$ rounds. In each round, identifying the active set $A$ incurs a linear scan cost of $O(n)$. The subsequent diffusion updates iterate over active vertices, and for each $u\in A$ the work is proportional to $d(u)$. Thus the total update cost in one round is $O(\sum_{u\in A} d(u)) \le O(\sum_{u\in V} d(u)) = O(m)$. Consequently, each round takes $O(n+m)$ time, yielding a total time complexity of $O(T(n+m))$. The space usage is $O(n+m)$, dominated by the graph storage and the two auxiliary energy functions $f$ and $f'$.
\end{proof}

\subsection{Quasi-Clique Extraction}

The procedure (Algorithm~\ref{alg:extractQC}) converts the energy function $f$ into a $\gamma$-quasi-clique by first ranking active vertices by energy, then selecting an initial candidate via a conductance sweep cut over energy-ranked prefixes, and finally refining the candidate to satisfy the density threshold $\gamma$.
Specifically, the sweep cut selects the minimum-conductance prefix to turn the ranking into an explicit candidate region. The refinement step then enforces feasibility under $\gamma$ through pruning and greedy expansion.

\begin{algorithm}[htb]
    \small
    \caption{\textsc{ExtractQuasiClique}$(G, f, \gamma, \theta)$}
    \label{alg:extractQC}
    \KwIn{Graph $G=(V,E)$, energy function $f$, density threshold $\gamma$, activation threshold $\theta$}
    \KwOut{A vertex set $S$ that induces a $\gamma$-quasi-clique}
    
    $A \gets \{v \in V \mid f(v) > \theta\}$\;
    $P \gets$ vertices of $A$ sorted by decreasing $f$\;
    \lIf{$|P| < 2$}{\Return{$\emptyset$}}
    
    $i^* \gets \arg\min_{i \in \{1, \dots, |P|-1\}} \Phi(\{v_1, \dots, v_i\})$\;
    $S \gets \{v_1, \dots, v_{i^*}\}$\;
    
    \While{$|S| > 1$ and $\eta(G[S]) < \gamma$}{
        $u \gets \arg\min_{x \in S} d_S(x)$\;
        $S \gets S \setminus \{u\}$\;
    }

    \lIf{$|S| < 2$}{\Return{$\emptyset$}}
    
    $\textit{changed} \gets \textbf{true}$\;
    \While{\textit{changed}}{
        $\textit{changed} \gets \textbf{false}$\;
        $R \gets$ vertices of $P \setminus S$, sorted by connections to $S$\;
        \ForEach{$v \in R$}{
            \If{$\eta(G[S \cup \{v\}]) \ge \gamma$}{
                $S \gets S \cup \{v\}$\;
                $\textit{changed} \gets \textbf{true}$\;
                \textbf{break}\;
            }
        }
    }
    \Return{$S$}
\end{algorithm}

First, the algorithm collects active vertices $A$ and sorts them by decreasing energy to form an ordered list $P$ (lines~1--2). If $|P|<2$, it returns $\emptyset$ (line~3).
This ranking is important, as it places vertices that are more likely to belong to a cohesive region near the front of the sequence.

Next, to obtain an initial candidate set, the algorithm performs a sweep cut~\cite{andersen2006local} on the energy-ranked list $P$ (lines~4--5).
Formally, let $P=(v_1,\dots,v_k)$. The procedure evaluates the conductance of each prefix $S_i=\{v_1,\dots,v_i\}$ for $1\le i\le k-1$, and selects the minimum-conductance prefix as the initial set $S$. 
Note that we evaluate prefixes up to $k-1$ to avoid the degenerate case $S=V$ (or $S=P$ here), for which $\Phi(S)$ is undefined as the denominator becomes zero.
In this way, the diffusion ranking is converted into a vertex set whose boundary is identified by the conductance metric.
Since $P$ is induced by diffusion scores, low-conductance prefixes are natural candidates for subsequent refinement.

Finally, the resulting initial set $S$ undergoes a refinement procedure (lines~6--18). The refinement begins with a pruning loop (lines~6--8). 
If the pruning reduces $S$ to a size smaller than $2$, the procedure returns an empty set (line~9), since a $\gamma$-quasi-clique requires at least two vertices.
If the initial set $S$ does not meet the density threshold, this loop iteratively removes the vertex with the minimum internal degree $d_S(u)$ until the set $S$ induces a valid $\gamma$-quasi-clique. 
Following this, an expansion phase seeks to enlarge the set while preserving feasibility (lines~10--18). 
In each pass, the remaining candidates $R=P\setminus S$ are re-sorted based on their connections to the current set $S$ (line~13). The algorithm then iterates through $R$ and greedily adds the first vertex $v$ that maintains the density constraint (lines~15--16). 
Upon a successful insertion, the current iteration over $R$ is terminated, since the connectivity ordering depends on the updated set $S$. The process repeats until no further insertion is possible without violating $\gamma$.

\begin{theorem}[Complexity of Algorithm~\ref{alg:extractQC}] 
Let $k=|A|$. \textsc{ExtractQuasiClique} runs in $O(k^2+\mathrm{vol}(A))$ time and uses $O(k)$ space.
\end{theorem}

\begin{proof}
Sorting the $k$ active vertices takes $O(k \log k)$. 
The sweep cut linearly scans the sorted list; computing the conductance of all prefixes via incremental updates takes $O(\mathrm{vol}(A))$ time.
The refinement stage entails a greedy optimization sequence. In the worst case, it performs $O(k)$ vertex additions or removals. By maintaining internal degrees incrementally, each update requires $O(k)$ operations, yielding an $O(k^2)$ bound for refinement. Since $k \log k \in O(k^2)$, the total time complexity is $O(k^2 + \mathrm{vol}(A))$.
The space usage is dominated by the ranked list and auxiliary structures for vertex membership and internal degrees, which use $O(k)$ space.
\end{proof}

\begin{theorem}[Complexity of EDQC]
Fix a source vertex $v$. Let $k=|A|$.
One iteration of EDQC with source $v$ runs in $O\!\big(T(k + \mathrm{vol}(A)) + k^2\big)$ time and uses $O(n+m)$ space.
The total time complexity for processing $B$ sources is the summation of this bound over the $B$ iterations.
In the degenerate worst case where $B=n$ and $k=\Theta(n)$ in every iteration, this bound recovers $O(n^3)$.
\end{theorem}

\begin{proof}
Consider a fixed source $v$.
The diffusion stage runs for $T$ rounds.
In our implementation, the active set is maintained incrementally, so each round updates only active vertices and scans their incident edges, incurring $O\!\left(T(k + \mathrm{vol}(A))\right)$ time.
The extraction stage sorts the $k$ active vertices in $O(k\log k)$. The subsequent sweep cut requires $O(\mathrm{vol}(A))$ time via incremental updates; since $\mathrm{vol}(A) \le T(k+\mathrm{vol}(A))$ for $T\ge 1$, this cost is subsumed by the diffusion stage.
The refinement stage performs a greedy optimization sequence; with incremental degree maintenance, its worst-case cost is bounded by $O(k^2)$.
Therefore, the total time complexity of one iteration is $O\!\big(T(k + \mathrm{vol}(A)) + k^2\big)$; in the degenerate worst case where $k=\Theta(n)$ and $B=n$, the total time over all sources becomes $O(n^3)$.
The space usage is $O(n+m)$, dominated by storing $G$ and two reusable energy functions $f$ and $f'$.
\end{proof}

\section{Experimental Evaluation}
\label{sec:experiments}

We conducted a comprehensive experimental evaluation of EDQC. We report solution quality, stability, and runtime under instance-specific density thresholds, together with statistical tests, ablations, and additional analyses on robustness across $\gamma$ and the energy--density premise. 

\subsection{Experimental Setup} \label{sec:setup}

All algorithms were implemented in C++ and compiled with g++ 9.4.0 using the ``-O3'' optimization flag. Experiments were run on a machine with an AMD EPYC 7H12 CPU (1.5 GHz base frequency), 128 GB of RAM, and Ubuntu 20.04.
Baseline parameters followed the original papers. For EDQC, we fixed $T=2$ and $\theta=10^{-4}$ (chosen based on the sensitivity study in Section~\ref{sec:sensitivity}) across all datasets.

Due to inherent randomness, all non-deterministic methods (all except NBSim) were executed 10 times with random seeds from 1 to 10; NBSim, being deterministic, was run once. For the non-deterministic methods, we report their average quasi-clique size and standard deviation (STD) over the 10 runs, presented in the format ``average size \(\pm\) STD'', where the standard deviation is reported only if it is non-zero. 
All runs were subject to a 60-second timeout. If an algorithm failed to find any valid $\gamma$-quasi-clique in all 10 runs for an instance, we mark its result as ``N/A''. The 60-second timeout reflects common practice in recent time-bounded heuristic studies~\cite{lin2017reduction,zheng2023farsighted}. We report wall-clock time excluding file I/O, and run all methods in a single thread.

\subsection{Baseline Methods}
We compared EDQC against four state-of-the-art algorithms:
\begin{itemize}
    \item \textbf{NBSim/FastNBSim}~\cite{pang2024similarity}: Similarity-based heuristics that expand subgraphs based on neighborhood overlap. FastNBSim accelerates NBSim via MinHash-based approximation.
    \item \textbf{NuQClqF1/NuQClqR1}~\cite{liu2024optimization}: Advanced variants of NuQClq~\cite{chen2021nuqclq}. These algorithms employ information feedback models that integrate historical solutions using fitness-weighted scores to guide local search.
\end{itemize}

\subsection{Datasets}
We evaluated all algorithms on 75 real-world datasets from SNAP~\cite{snapnets} and the Network Repository~\cite{nr}. The collection covers diverse domains, including web, social, and communication graphs. The datasets span a wide spectrum of scales, with vertex counts ($n$) ranging from 62 to over 16.7 million, and edge densities spanning from $8.41 \times 10^{-2}$ down to as low as $1.87 \times 10^{-7}$.

\subsection{Main Results} \label{sec:main-results}

This section reports the main results on solution quality, stability, and runtime under instance-specific density thresholds induced by NBSim and FastNBSim. A key challenge in this comparison is that NBSim and FastNBSim do not take a density threshold $\gamma$ as input, whereas EDQC, NuQClqF1, and NuQClqR1 do. To enable a comparable evaluation across these methods, we adopt two evaluation settings, following the protocol of Pang et al.~\cite{pang2024similarity}.

\begin{table}[htb]
\centering
\caption{Quasi-clique sizes with the threshold $\gamma$ set by NBSim's output.}
\resizebox{\columnwidth}{!}{
\begin{tabular}{lcrrrr}
\toprule
Instance & $\gamma$ & NBSim & NuQClqF1 & NuQClqR1 & EDQC \\
\midrule
rt-retweet & 1 & 3 & \textbf{4} & \textbf{4} & \textbf{4} \\
web-polblogs & 1 & 5 & \textbf{9} & \textbf{9} & \textbf{9} \\
email-Eu-core & 1 & 12 & \textbf{18} & \textbf{18} & \textbf{18} \\
ego-facebook & 0.99 & 71 & \textbf{92} & \textbf{92} & \textbf{92} \\
ca-Erdos992 & 1 & 6 & \textbf{8} & \textbf{8} & \textbf{8} \\
socfb-CMU & 0.98 & 20 & \textbf{53} & \textbf{53} & \textbf{53} \\
ech-WHOIS & 0.98 & 57 & \textbf{79} & \textbf{79} & \textbf{79} \\
web-indochina-2004 & 1 & \textbf{50} & 49 $\pm$ {\footnotesize 3.16} & \textbf{50} & \textbf{50} \\
socfb-UCSB37 & 0.99 & 40 & \textbf{64} & \textbf{64} & \textbf{64} \\
socfb-UConn & 0.99 & 22 & \textbf{58} & \textbf{58} & \textbf{58} \\
socfb-UCLA & 0.79 & 19 & \textbf{96} & \textbf{96} & \textbf{96} \\
socfb-Berkeley13 & 1 & 21 & \textbf{42} & \textbf{42} & \textbf{42} \\
socfb-Wisconsin87 & 0.98 & 12 & \textbf{46} & \textbf{46} & \textbf{46} \\
soc-epinions & 1 & 10 & \textbf{16} & \textbf{16} & \textbf{16} \\
Cit-HepTh & 1 & 12 & 22.9 $\pm$ {\footnotesize 0.32} & \textbf{23} & \textbf{23} \\
socfb-Indiana & 0.85 & 27 & \textbf{94} & \textbf{94} & \textbf{94} \\
socfb-UIllinois & 1 & 14 & \textbf{57} & \textbf{57} & \textbf{57} \\
Cit-HepPh & 0.98 & 14 & \textbf{23} & \textbf{23} & \textbf{23} \\
socfb-UF & 0.99 & 24 & \textbf{67} & 66.8 $\pm$ {\footnotesize 0.63} & \textbf{67} \\
Email-Enron & 0.97 & 10 & \textbf{25} & \textbf{25} & \textbf{25} \\
socfb-Penn94 & 0.98 & 14 & \textbf{56} & 55.3 $\pm$ {\footnotesize 2.21} & \textbf{56} \\
soc-brightkite & 0.99 & 36 & \textbf{45} & \textbf{45} & \textbf{45} \\
socfb-OR & 0.6 & 19 & \textbf{108} & \textbf{108} & \textbf{108} \\
soc-slashdot & 1 & 12 & \textbf{26} & \textbf{26} & \textbf{26} \\
witter\_combined & 0.98 & 79 & 89.3 $\pm$ {\footnotesize 11.7} & 89.3 $\pm$ {\footnotesize 11.7} & \textbf{93} \\
G\_n\_pin\_pout & 1 & 3 & \textbf{4} & \textbf{4} & \textbf{4} \\
yahoo-msg & 0.99 & 22 & 20.6 $\pm$ {\footnotesize 2.07} & 20.6 $\pm$ {\footnotesize 2.01} & \textbf{23} \\
soc-douban & 1 & 5 & \textbf{11} & \textbf{11} & \textbf{11} \\
wave & 1 & 5 & \textbf{6} & \textbf{6} & \textbf{6} \\
Loc-Gowalla & 0.99 & \textbf{31} & 23.8 $\pm$ {\footnotesize 2.53} & 23 & \textbf{31} \\
soc-gowalla & 0.99 & \textbf{31} & 23 & 23.8 $\pm$ {\footnotesize 2.53} & \textbf{31} \\
web-Stanford & 0.99 & 67 & 64.2 $\pm$ {\footnotesize 3.43} & 59.9 $\pm$ {\footnotesize 4.12} & \textbf{67.5} $\pm$ {\footnotesize 0.71} \\
cnr-2000 & 0.99 & \textbf{89} & 84.8 $\pm$ {\footnotesize 3.61} & 81.8 $\pm$ {\footnotesize 0.63} & \textbf{89} \\
web-NotreDame & 1 & \textbf{155} & 154.3 $\pm$ {\footnotesize 0.48} & 154.2 $\pm$ {\footnotesize 0.42} & \textbf{155} \\
ca-MathSciNet & 1 & \textbf{25} & 24.6 $\pm$ {\footnotesize 0.84} & 24.7 $\pm$ {\footnotesize 0.67} & \textbf{25} \\
coPapersDBLP & 1 & \textbf{337} & 333.5 $\pm$ {\footnotesize 4.12} & 333.5 $\pm$ {\footnotesize 4.12} & \textbf{337} \\
auto & 1 & 6 & \textbf{7} & \textbf{7} & \textbf{7} \\
web-it-2004 & 1 & \textbf{432} & 431.8 $\pm$ {\footnotesize 0.42} & \textbf{432} & \textbf{432} \\
soc-delicious & 0.58 & 22 & \textbf{80} & \textbf{80} & \textbf{80} \\
ca-coauthors-dblp & 1 & \textbf{337} & 333.5 $\pm$ {\footnotesize 4.12} & 333.5 $\pm$ {\footnotesize 4.12} & \textbf{337} \\
eu-2005 & 0.99 & 405 & 191.3 $\pm$ {\footnotesize 115.84} & 127.2 $\pm$ {\footnotesize 26.92} & \textbf{405.1} $\pm$ {\footnotesize 0.32} \\
web-Google & 0.99 & \textbf{48} & 39.2 $\pm$ {\footnotesize 7.86} & 39.5 $\pm$ {\footnotesize 7.41} & \textbf{48} \\
rgg\_n\_2\_20\_s0 & 0.57 & 17 & \textbf{41} & \textbf{41} & 40.2 $\pm$ {\footnotesize 0.79} \\
rgg\_n\_2\_21\_s0 & 1 & 18 & 18.1 $\pm$ {\footnotesize 0.57} & 18.1 $\pm$ {\footnotesize 0.57} & \textbf{19} \\
rgg\_n\_2\_22\_s0 & 1 & 19 & 18.8 $\pm$ {\footnotesize 0.79} & 18.6 $\pm$ {\footnotesize 0.7} & \textbf{20} \\
rgg\_n\_2\_23\_s0 & 0.99 & 19 & 18.3 $\pm$ {\footnotesize 0.48} & 19 $\pm$ {\footnotesize 1.41} & \textbf{22} \\
rgg\_n\_2\_24\_s0 & 0.98 & 20 & 19.7 $\pm$ {\footnotesize 0.95} & 18.7 $\pm$ {\footnotesize 0.48} & \textbf{23} \\
\bottomrule
\end{tabular}
}
\label{tab:1}
\end{table}

In the first setting, NBSim is run once per instance, and the density of its output quasi-clique is used as the target threshold $\gamma$ for EDQC, NuQClqF1, and NuQClqR1. 
In the second setting, FastNBSim is run 10 times following the protocol in Section~\ref{sec:setup}; the density threshold $\gamma$ is then set to the maximum density achieved over these runs, which reflects the strongest cohesion level FastNBSim can reach on the instance, and EDQC, NuQClqF1, and NuQClqR1 are evaluated under this common threshold. The detailed quasi-clique sizes in these two settings are reported in Tables~\ref{tab:1} and~\ref{tab:2}; to save space, instances where all algorithms obtain quasi-cliques of identical average size are omitted.

\begin{table}[!t]
\centering
\caption{Quasi-clique sizes with $\gamma$ set by FastNBSim's max density over 10 runs. For FastNBSim, stats are on successful runs only (rate shown in parentheses).}
\resizebox{\columnwidth}{!}{
\begin{tabular}{lcrrrr}
\toprule
Instance & $\gamma$ & FastNBSim & NuQClqF1 & NuQClqR1 & EDQC \\
\midrule
soc-dolphins & 0.83 & 4 \footnotesize{(4/10)} & \textbf{6} & \textbf{6} & \textbf{6} \\
rt-retweet & 1 & 3 \footnotesize{(7/10)} & \textbf{4} & \textbf{4} & \textbf{4} \\
ca-netscience & 1 & 8.1 $\pm$ {\footnotesize 0.88} \footnotesize{(10/10)} & \textbf{9} & \textbf{9} & \textbf{9} \\
web-polblogs & 0.93 & 6 \footnotesize{(2/10)} & \textbf{13} & \textbf{13} & \textbf{13} \\
rt-twitter-copen & 1 & 3.6 $\pm$ {\footnotesize 0.53} \footnotesize{(7/10)} & \textbf{4} & \textbf{4} & \textbf{4} \\
email-Eu-core & 0.94 & 21 \footnotesize{(1/10)} & \textbf{29} & \textbf{29} & \textbf{29} \\
web-edu & 1 & 27.4 $\pm$ {\footnotesize 3.1} \footnotesize{(10/10)} & \textbf{30} & \textbf{30} & \textbf{30} \\
ego-facebook & 0.96 & 102 \footnotesize{(1/10)} & \textbf{123} & \textbf{123} & \textbf{123} \\
ca-GrQc & 1 & 37.8 $\pm$ {\footnotesize 3.2} \footnotesize{(4/10)} & \textbf{44} & \textbf{44} & \textbf{44} \\
web-spam & 1 & 15.5 $\pm$ {\footnotesize 0.71} \footnotesize{(2/10)} & \textbf{20} & \textbf{20} & \textbf{20} \\
ca-Erdos992 & 1 & 5 \footnotesize{(6/10)} & \textbf{8} & \textbf{8} & \textbf{8} \\
socfb-CMU & 0.97 & 28 \footnotesize{(1/10)} & \textbf{56} & \textbf{56} & \textbf{56} \\
ech-WHOIS & 0.96 & 73 \footnotesize{(1/10)} & \textbf{91} & \textbf{91} & \textbf{91} \\
ca-HepPh & 1 & 202.7 $\pm$ {\footnotesize 24.04} \footnotesize{(10/10)} & \textbf{239} & \textbf{239} & \textbf{239} \\
web-indochina-2004 & 1 & 49.9 $\pm$ {\footnotesize 0.32} \footnotesize{(10/10)} & 49 $\pm$ {\footnotesize 3.16} & \textbf{50} & \textbf{50} \\
web-BerkStan & 1 & 26.7 $\pm$ {\footnotesize 2.83} \footnotesize{(10/10)} & \textbf{29} & \textbf{29} & \textbf{29} \\
socfb-UCSB37 & 0.98 & 47 \footnotesize{(1/10)} & \textbf{70} & \textbf{70} & \textbf{70} \\
web-webbase-2001 & 1 & 32.9 $\pm$ {\footnotesize 0.32} \footnotesize{(10/10)} & \textbf{33} & \textbf{33} & \textbf{33} \\
socfb-UConn & 0.97 & 34 \footnotesize{(1/10)} & \textbf{65} & \textbf{65} & \textbf{65} \\
ca-AstroPh & 1 & 51.8 $\pm$ {\footnotesize 3.12} \footnotesize{(10/10)} & \textbf{57} & \textbf{57} & \textbf{57} \\
socfb-UCLA & 0.98 & 49 \footnotesize{(1/10)} & \textbf{63} & \textbf{63} & \textbf{63} \\
ca-CondMat & 1 & 22.63 $\pm$ {\footnotesize 2.83} \footnotesize{(8/10)} & \textbf{26} & \textbf{26} & \textbf{26} \\
socfb-Berkeley13 & 0.95 & 29 \footnotesize{(1/10)} & \textbf{62} & \textbf{62} & \textbf{62} \\
socfb-Wisconsin87 & 0.91 & 34 \footnotesize{(1/10)} & \textbf{65} & \textbf{65} & \textbf{65} \\
soc-epinions & 0.95 & 16 \footnotesize{(1/10)} & \textbf{21} & \textbf{21} & \textbf{21} \\
Cit-HepTh & 0.92 & 26 \footnotesize{(1/10)} & \textbf{41} & \textbf{41} & \textbf{41} \\
socfb-Indiana & 0.95 & 40 \footnotesize{(1/10)} & \textbf{71} & 70.8 $\pm$ {\footnotesize 0.63} & \textbf{71} \\
socfb-UIllinois & 0.94 & 48 \footnotesize{(1/10)} & \textbf{92} & \textbf{92} & \textbf{92} \\
Cit-HepPh & 0.98 & 20 \footnotesize{(1/10)} & \textbf{23} & \textbf{23} & \textbf{23} \\
socfb-UF & 0.94 & 69 \footnotesize{(1/10)} & \textbf{93} & \textbf{93} & \textbf{93} \\
Email-Enron & 0.96 & 13 \footnotesize{(1/10)} & \textbf{26} & \textbf{26} & \textbf{26} \\
socfb-Penn94 & 0.94 & 42 \footnotesize{(1/10)} & 63.4 $\pm$ {\footnotesize 3.37} & 63.3 $\pm$ {\footnotesize 3.59} & \textbf{65} \\
soc-brightkite & 0.97 & 43 \footnotesize{(1/10)} & \textbf{53} & \textbf{53} & \textbf{53} \\
socfb-OR & 0.92 & 24 \footnotesize{(1/10)} & 50.9 $\pm$ {\footnotesize 0.32} & 50.9 $\pm$ {\footnotesize 0.32} & \textbf{51} \\
soc-slashdot & 0.97 & 10 \footnotesize{(1/10)} & \textbf{35} & \textbf{35} & \textbf{35} \\
wordnet-words & 1 & 29.2 $\pm$ {\footnotesize 2.71} \footnotesize{(6/10)} & \textbf{32} & \textbf{32} & \textbf{32} \\
witter\_combined & 0.97 & 87 \footnotesize{(1/10)} & \textbf{99} & \textbf{99} & \textbf{99} \\
G\_n\_pin\_pout & 0.5 & 4 \footnotesize{(3/10)} & 7.9 $\pm$ {\footnotesize 0.32} & 7.9 $\pm$ {\footnotesize 0.32} & \textbf{8} \\
yahoo-msg & 0.96 & 24 \footnotesize{(1/10)} & 23.4 $\pm$ {\footnotesize 1.84} & 23.9 $\pm$ {\footnotesize 1.1} & \textbf{25} \\
web-sk-2005 & 0.99 & 83 \footnotesize{(3/10)} & \textbf{84} & \textbf{84} & \textbf{84} \\
soc-douban & 0.72 & 9 \footnotesize{(1/10)} & \textbf{23} & \textbf{23} & \textbf{23} \\
wave & 0.71 & 8 \footnotesize{(3/10)} & \textbf{14} & \textbf{14} & \textbf{14} \\
web-arabic-2005 & 1 & 90.7 $\pm$ {\footnotesize 5.36} \footnotesize{(10/10)} & \textbf{102} & \textbf{102} & \textbf{102} \\
Loc-Gowalla & 1 & 23.5 $\pm$ {\footnotesize 0.71} \footnotesize{(2/10)} & 21 & 21 & \textbf{29} \\
soc-gowalla & 0.99 & 25 \footnotesize{(1/10)} & 23 & 23.8 $\pm$ {\footnotesize 2.53} & \textbf{31} \\
ca-dblp-2010 & 1 & 70.6 $\pm$ {\footnotesize 4.86} \footnotesize{(10/10)} & \textbf{75} & \textbf{75} & \textbf{75} \\
ca-citeseer & 1 & 81.6 $\pm$ {\footnotesize 5.36} \footnotesize{(10/10)} & \textbf{87} & \textbf{87} & \textbf{87} \\
coAuthorsCiteseer & 1 & 81.6 $\pm$ {\footnotesize 5.36} \footnotesize{(10/10)} & \textbf{87} & \textbf{87} & \textbf{87} \\
web-Stanford & 0.99 & 62 \footnotesize{(1/10)} & 64.2 $\pm$ {\footnotesize 3.43} & 59.9 $\pm$ {\footnotesize 4.12} & \textbf{67.5} $\pm$ {\footnotesize 0.71} \\
coAuthorsDBLP & 1 & 103 $\pm$ {\footnotesize 9.7} \footnotesize{(10/10)} & \textbf{115} & \textbf{115} & \textbf{115} \\
ca-dblp-2012 & 1 & 104 $\pm$ {\footnotesize 9.52} \footnotesize{(10/10)} & \textbf{114} & \textbf{114} & \textbf{114} \\
com-dblp & 1 & 101.4 $\pm$ {\footnotesize 5.76} \footnotesize{(10/10)} & \textbf{114} & \textbf{114} & \textbf{114} \\
cnr-2000 & 1 & \textbf{84} \footnotesize{(1/10)} & 81.9 $\pm$ {\footnotesize 1.45} & 81.8 $\pm$ {\footnotesize 1.55} & \textbf{84} \\
web-NotreDame & 1 & 153.5 $\pm$ {\footnotesize 1.43} \footnotesize{(10/10)} & 154.3 $\pm$ {\footnotesize 0.48} & 154.2 $\pm$ {\footnotesize 0.42} & \textbf{155} \\
ca-MathSciNet & 1 & 23.8 $\pm$ {\footnotesize 0.79} \footnotesize{(10/10)} & 24.6 $\pm$ {\footnotesize 0.84} & 24.5 $\pm$ {\footnotesize 0.85} & \textbf{25} \\
coPapersDBLP & 1 & 316.9 $\pm$ {\footnotesize 23} \footnotesize{(10/10)} & 333 $\pm$ {\footnotesize 3.94} & 333 $\pm$ {\footnotesize 3.94} & \textbf{337} \\
auto & 0.77 & 9.5 $\pm$ {\footnotesize 0.71} \footnotesize{(2/10)} & \textbf{11} & \textbf{11} & \textbf{11} \\
web-it-2004 & 1 & 431.7 $\pm$ {\footnotesize 0.48} \footnotesize{(10/10)} & 431.8 $\pm$ {\footnotesize 0.42} & \textbf{432} & \textbf{432} \\
soc-delicious & 0.88 & 30 \footnotesize{(1/10)} & 31.4 $\pm$ {\footnotesize 3.86} & 31.4 $\pm$ {\footnotesize 3.86} & \textbf{37} \\
ca-coauthors-dblp & 1 & 316.9 $\pm$ {\footnotesize 23} \footnotesize{(10/10)} & 333.5 $\pm$ {\footnotesize 4.12} & 330.2 $\pm$ {\footnotesize 10.01} & \textbf{337} \\
soc-digg & 0.9 & 76 \footnotesize{(1/10)} & 104 $\pm$ {\footnotesize 13.7} & 98.8 $\pm$ {\footnotesize 12.56} & \textbf{116.9} $\pm$ {\footnotesize 0.32} \\
eu-2005 & 0.99 & 387 \footnotesize{(1/10)} & 191.3 $\pm$ {\footnotesize 115.84} & 127.2 $\pm$ {\footnotesize 26.92} & \textbf{405.1} $\pm$ {\footnotesize 0.32} \\
web-Google & 1 & 40.7 $\pm$ {\footnotesize 3.21} \footnotesize{(3/10)} & 33.7 $\pm$ {\footnotesize 2.63} & 37.1 $\pm$ {\footnotesize 6.08} & \textbf{44} \\
soc-pokec & 0.99 & \textbf{31} \footnotesize{(1/10)} & 24.6 $\pm$ {\footnotesize 2.37} & 23.8 $\pm$ {\footnotesize 1.55} & 25 \\
rgg\_n\_2\_22\_s0 & 0.98 & 19 \footnotesize{(1/10)} & 20.1 $\pm$ {\footnotesize 0.99} & 20 $\pm$ {\footnotesize 0.94} & \textbf{21} \\
rgg\_n\_2\_23\_s0 & 0.99 & 21 \footnotesize{(1/10)} & 18.3 $\pm$ {\footnotesize 0.48} & 19 $\pm$ {\footnotesize 1.41} & \textbf{22} \\
hugetrace-00010 & 0.5 & 4 \footnotesize{(10/10)} & \footnotesize{N/A} & \footnotesize{N/A} & \textbf{5} \\
hugetrace-00020 & 0.5 & 4 \footnotesize{(10/10)} & \footnotesize{N/A} & \footnotesize{N/A} & \textbf{5} \\
rgg\_n\_2\_24\_s0 & 0.95 & 20 \footnotesize{(1/10)} & 20.6 $\pm$ {\footnotesize 1.26} & 20.9 $\pm$ {\footnotesize 1.37} & \textbf{25} \\
\bottomrule
\end{tabular}
}
\label{tab:2}
\end{table}

\paragraph{Effectiveness and Stability.}
The results in Table~\ref{tab:1} and Table~\ref{tab:2} show the superior effectiveness and stability of EDQC. A quantitative summary highlights this advantage. In the first setting (Table~\ref{tab:1}), across the 47 instances where results differed, EDQC achieves or ties for the best average size in 46 instances. This compares favorably to the other methods, where NuQClqR1 achieved or tied for the best size in 29 instances, followed by NuQClqF1 (28 instances) and NBSim (10 instances). Similarly, in the second setting (Table~\ref{tab:2}), EDQC achieves or ties for the best size in 68 out of 69 instances. For comparison, the counts for the other algorithms were: NuQClqR1 (46 instances), NuQClqF1 (45 instances), and FastNBSim (2 instances).

Beyond average size, EDQC demonstrates exceptional stability. It consistently reports a standard deviation close to zero (e.g., $0.32$ on the large \texttt{eu-2005} graph under the induced $\gamma$), even on complex instances. This is in stark contrast to the NuQClq variants, which frequently exhibit high variance on the same instances (e.g., a STD of $115.84$ for NuQClqF1 on \texttt{eu-2005}). Furthermore, the results highlight the unreliability of the similarity-based methods. FastNBSim, for instance, often fails to achieve its own best-found density (succeeding only 1/10 times on many graphs). This lack of explicit density control makes them less suitable for applications requiring highly cohesive structures, a task where EDQC's reliability is a distinct advantage.

\begin{figure}[htb]
\centering
\includegraphics[width=0.7\linewidth]{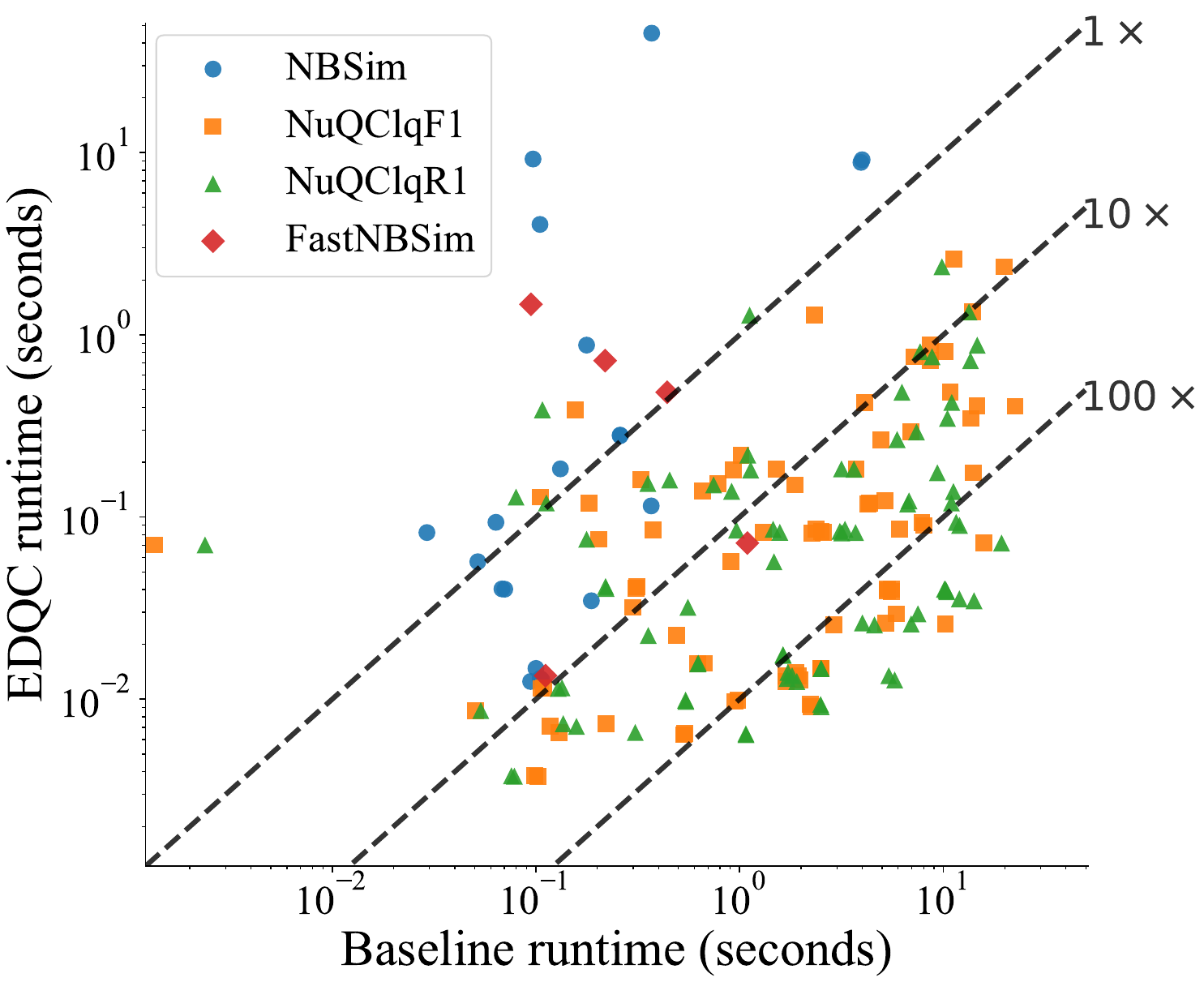} 
\caption{Runtime comparison of EDQC against baselines on a log-log scale.}
\label{fig:runtime}
\end{figure}

\paragraph{Efficiency.}
The efficiency of EDQC is compared against baselines in Figure~\ref{fig:runtime}, where, following prior work~\cite{sunnumds}, instances where the runtime of both EDQC and the corresponding baseline is below 0.1 seconds are excluded. 
On this log-log plot, points falling below the $1\times$, $10\times$, and $100\times$ lines indicate that EDQC is faster than the baseline by the corresponding factor or more. 
The results indicate that EDQC is substantially more efficient than the local search methods, NuQClqF1 and NuQClqR1, with many points falling below the $10\times$ line, signifying a speedup of over an order of magnitude.
In comparison to the similarity-based heuristics (NBSim and FastNBSim), EDQC's runtime remains competitive, as most of their corresponding points cluster around the $1\times$ line. Overall, this demonstrates that EDQC strikes a competitive balance between achieving state-of-the-art solution quality and maintaining high computational efficiency.

\subsection{Statistical Analysis}
\label{sec:statistical-analysis}

To assess statistical significance, we applied the Friedman test~\cite{friedman1937use} on the algorithm rankings across all 75 datasets. The null hypothesis that all algorithms perform equally was rejected in both settings ($p < 0.05$), allowing us to proceed with the post-hoc Nemenyi test~\cite{demvsar2006statistical} to identify pairwise performance differences.
On each dataset, we rank algorithms by the reported quasi-clique size under the same induced $\gamma$ (for randomized methods, the mean size over 10 runs). Ties receive averaged ranks. When a method is marked as ``N/A'' on a dataset, we treat it as the worst performance for ranking on that dataset.

\begin{figure}[t]
    \centering

    \begin{subfigure}{\linewidth}
        \centering
        \includegraphics[width=0.95\linewidth]{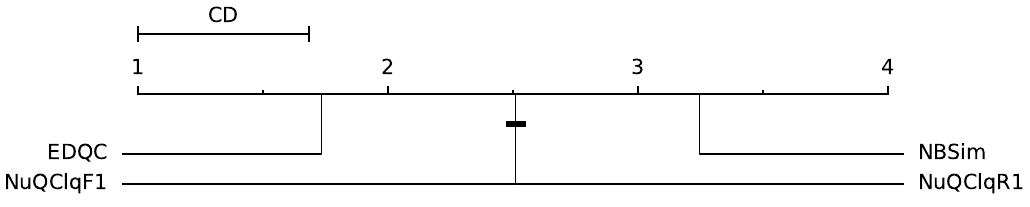}
        \caption{Comparison in the NBSim setting.}
        \label{fig:cd-nbsim}
    \end{subfigure}

    \vspace{1em} 

    \begin{subfigure}{\linewidth}
        \centering
        \includegraphics[width=0.95\linewidth]{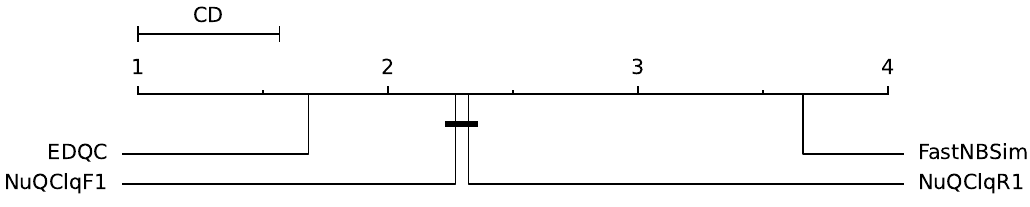}
        \caption{Comparison in the FastNBSim setting.}
        \label{fig:cd-fastnbsim}
    \end{subfigure}

    \caption{Critical Difference diagrams for algorithm ranks on 75 datasets (significance level 0.05). A horizontal bar connects algorithms with no statistically significant difference.}
    \label{fig:cd-plots}
\end{figure}

The results are visualized using Critical Difference (CD) diagrams in Figure~\ref{fig:cd-plots}. In these diagrams, algorithms are plotted on an axis according to their average rank across all datasets; a lower rank (further to the left) is better. Groups of algorithms whose performance is not statistically significantly different are connected by a horizontal bar, the length of which corresponds to the critical difference.

The top plot in Figure~\ref{fig:cd-plots} shows the comparison for the NBSim setting. EDQC achieves the best average rank, and this leading position is statistically significant. The post-hoc test reveals that EDQC's performance is significantly better than all three other baselines: NuQClqF1, NuQClqR1, and NBSim.
A similar trend is observed in the FastNBSim setting, shown in the bottom plot. Once again, EDQC obtains the best average rank. Its superiority is also confirmed to be statistically significant, as it significantly outperforms NuQClqF1, NuQClqR1, and FastNBSim.

In summary, the statistical tests provide strong evidence that EDQC achieves better average ranks than the compared baselines under both experimental settings at the 0.05 significance level.

\subsection{Analysis of EDQC Components}
\label{sec:ablation}

To isolate and verify the contribution of each key component, we conducted an ablation study comparing the full EDQC against five variants across 112 unique experimental settings (consolidating settings where both configurations yielded the same $\gamma$ value). These variants were designed to test the core aspects of our algorithm: the subgraph identification strategy (EDQC-1), the energy-guided ranking (EDQC-2, EDQC-3), the refinement phase (EDQC-4), and the nature of the diffusion process itself (EDQC-5). 

The results, summarized in Table~\ref{tab:ablation}, provide strong evidence that each component contributes to the overall performance of EDQC. The most dramatic performance degradation occurred with EDQC-5, which replaced our adaptive diffusion with a naive random process, and EDQC-1, which substituted conductance-based identification with a max-density heuristic. The full EDQC algorithm proved superior in 106 and 69 instances, respectively. These results underscore that our principled diffusion and extraction mechanisms form the foundational pillars of the algorithm's success. The importance of the energy-guided ranking was also clearly demonstrated; disrupting the descending energy order with a random (EDQC-2) or ascending (EDQC-3) ranking severely hampered performance, leading to 56 and 70 wins for the full EDQC. Finally, disabling the refinement phase (EDQC-4) also resulted in a noticeable performance drop in 16 instances, confirming that this final step is vital for optimizing the solution size.

Additionally, we confirmed the robustness of our synchronous design. As variants with a fixed processing order for active vertices produced identical results to the full EDQC, the order-independence guaranteed by the synchronous design was empirically verified.

\begin{table}[htb]
\centering
\caption{Comparison of the full EDQC with its five simplified variants across 112 experimental settings. \#bet and \#wor represent the number of instances where EDQC achieves better and worse quasi-clique sizes, respectively.}
\label{tab:ablation}
\resizebox{\columnwidth}{!}{
\begin{tabular}{llcc}
\toprule
\textbf{Variant} & \textbf{Description} & \textbf{\#bet} & \textbf{\#wor} \\
& & \small(EDQC > Variant) & \small(EDQC < Variant) \\
\midrule
EDQC-1 & No Conductance& 69 & 0 \\
EDQC-2 & Random Ranking & 56 & 1 \\
EDQC-3 & Ascending Ranking & 70 & 1 \\
EDQC-4 & No Refinement Phase & 16 & 1 \\
EDQC-5 & Naive Random Diffusion & 106 & 0 \\
\bottomrule
\end{tabular}
}
\end{table}

\subsection{Results under Fixed Density Thresholds}
\label{sec:fixed-gamma}

\begin{figure}[b]
\centering
\includegraphics[width=\linewidth]{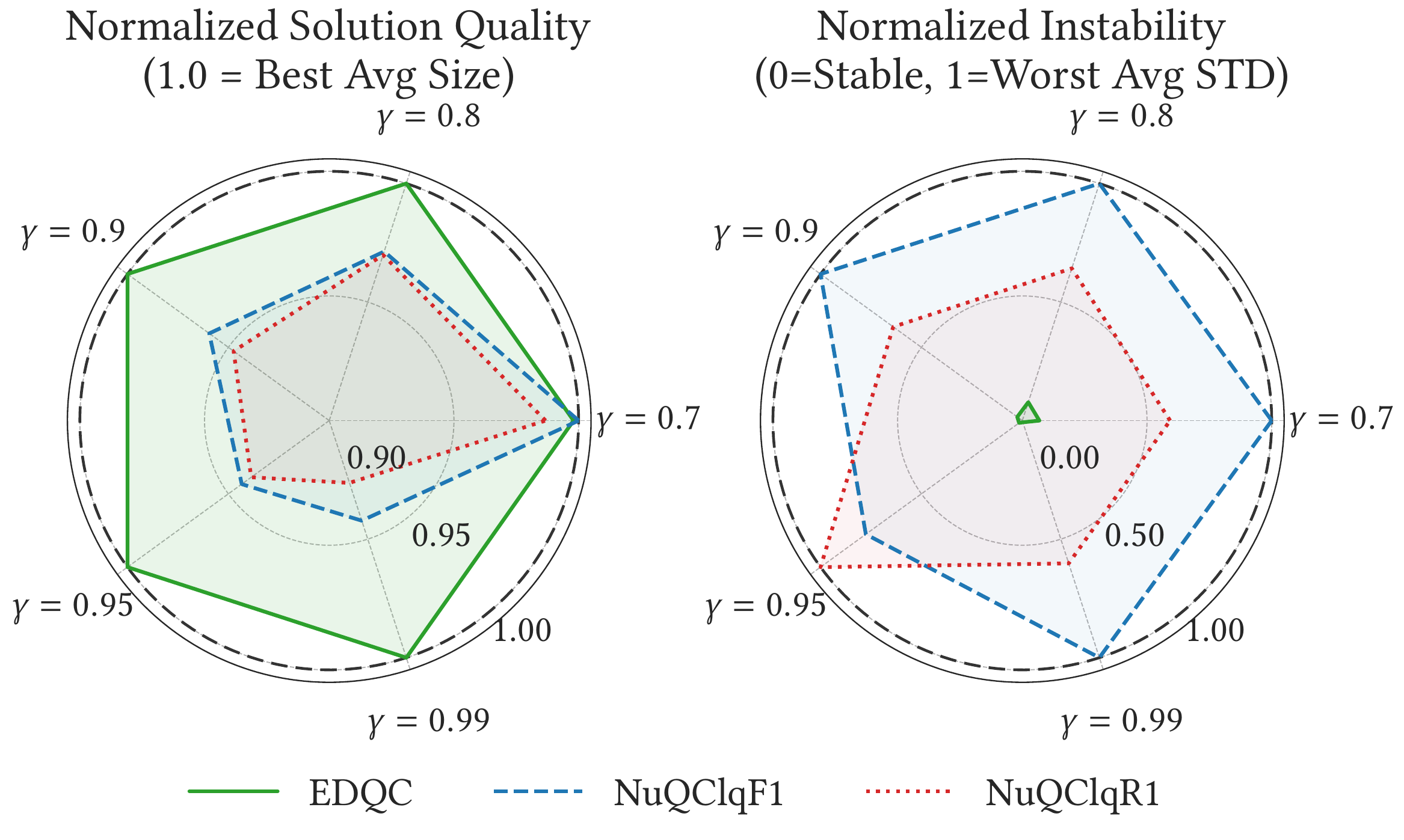} 
\caption{Fixed-$\gamma$ comparison of EDQC, NuQClqF1, and NuQClqR1 on 75 graphs for $\gamma\in\{0.7,0.8,0.9,0.95,0.99\}$.
Left: normalized solution quality (1.0 = best average quasi-clique size at that $\gamma$).
Right: normalized instability (0 = lowest, 1 = highest average standard deviation at that $\gamma$).}
\label{fig:fixed-gamma-radar}
\end{figure}

Beyond the instance-specific thresholds induced by NBSim and FastNBSim, we also examine how algorithms that accept an explicit density parameter behave across a shared set of fixed thresholds. Specifically, we compare EDQC, NuQClqF1, and NuQClqR1 on all 75 datasets under $\gamma \in \{0.7, 0.8, 0.9, 0.95, 0.99\}$. We exclude NBSim and FastNBSim here since they do not take $\gamma$ as input.

For each graph and each $\gamma \in \{0.7, 0.8, 0.9, 0.95, 0.99\}$, the three algorithms are run under the same protocol as in Section~\ref{sec:setup}. For each dataset and each (algorithm, $\gamma$), we compute the mean and standard deviation (STD) of the quasi-clique size over 10 runs, and then average these quantities over the 75 datasets. If an algorithm fails to find any valid $\gamma$-quasi-clique on a dataset, we treat it as the worst performance for aggregation.

The left panel of Figure~\ref{fig:fixed-gamma-radar} shows a radar plot of the normalized solution quality, where for each $\gamma$ the best average size among the three algorithms is scaled to $1.0$ (computed per $\gamma$ across the three algorithms). The right panel reports the normalized instability, where for each $\gamma$ the lowest average STD is mapped to $0$ and the highest to $1$ (again per $\gamma$). We use a radar plot to visualize the performance profile across multiple $\gamma$ values.

\begin{figure*}[htb]
\centering
\includegraphics[width=0.8\linewidth]{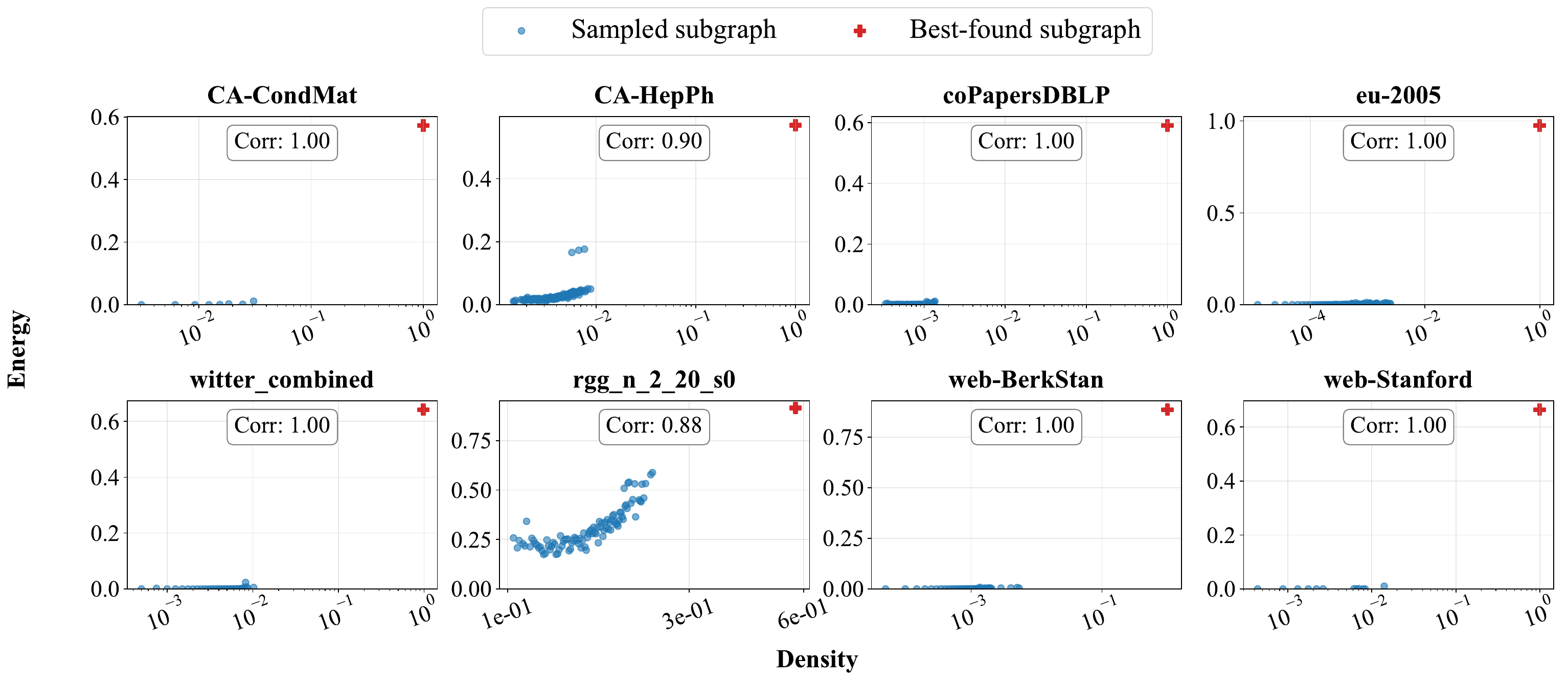}
\caption{Subgraph density vs. total retained energy on eight representative datasets. Each blue dot represents one of 1,000 randomly sampled subgraphs of a fixed size; the red marks the quasi-clique of the same size found by EDQC.}
\label{fig:energy}
\end{figure*}

The plots reveal a consistent pattern. On the solution-quality radar (left), EDQC stays close to $1.0$ across all density thresholds, indicating that it attains the best (or tied-best) average size at each $\gamma$. Moreover, the gap in normalized size tends to widen as $\gamma$ increases from $0.7$ to $0.99$, suggesting that EDQC is particularly strong when stricter cohesion is required. On the instability radar (right), EDQC remains near $0$ across all densities, reflecting consistently low variability, while NuQClqF1 shows higher normalized instability and NuQClqR1 lies in between. These fixed-threshold results provide additional evidence that EDQC delivers both high solution quality and low run-to-run variability over a broad range of density requirements.

\subsection{Structural characterization on \texttt{web-Google}}
\label{sec:struct-webgoogle}

We further examine the structural properties of the $\gamma$-quasi-cliques returned by EDQC and NuQClqF1/R1 on the large web graph \texttt{web-Google} at $\gamma = 0.9$. Table~\ref{tab:webgoogle-struct} reports the mean and standard deviation over 10 runs. EDQC finds substantially larger groups ($60.0 \pm 0.0$ nodes vs.\ $43.8 \pm 8.4$ and $41.2 \pm 6.5$) while maintaining similarly high internal density. At the same time, its conductance is much lower and far more stable ($0.101 \pm 0.003$ vs.\ $0.570 \pm 0.305$ and $0.669 \pm 0.235$), indicating that EDQC consistently identifies larger and better isolated cohesive groups. The clustering coefficients are very high and comparable ($\approx 0.92$–$0.93$), suggesting similar internal triangle structure across methods.

\begin{table}[t]
    \centering
    \caption{Structural properties of $\gamma$-quasi-cliques on \texttt{web-Google} at $\gamma = 0.9$. Mean $\pm$ standard deviation over 10 runs.}
    \label{tab:webgoogle-struct}
    \resizebox{\linewidth}{!}{
    \begin{tabular}{lcccc}
        \toprule
        Method & $|S|$ & Density & $\Phi(S)$ & Clust.\ coeff. \\
        \midrule
        EDQC      & $60.0$   & $0.904 \pm 0.002$ & $0.101 \pm 0.003$ & $0.921 \pm 0.001$ \\
        NuQClqF1  & $43.8 \pm 8.4$   & $0.911 \pm 0.010$ & $0.570 \pm 0.305$ & $0.934 \pm 0.015$ \\
        NuQClqR1  & $41.2 \pm 6.5$   & $0.907 \pm 0.004$ & $0.669 \pm 0.235$ & $0.929 \pm 0.015$ \\
        \bottomrule
    \end{tabular}
    }
\end{table}

\subsection{Empirical Validation of the Energy-Density Premise}
\label{sec:correlation}

To empirically validate our algorithm's core premise that energy concentrates in dense regions, we analyzed the relationship between subgraph density and energy retention on eight representative datasets. For each dataset, we compared the quasi-clique found by EDQC against 1,000 randomly sampled subgraphs of the same size. The results, visualized in Figure~\ref{fig:energy}, reveal a strong positive correlation between density and retained energy, confirmed by high Pearson correlation coefficients (typically $>0.90$). More importantly, the quasi-clique discovered by EDQC (red marks) consistently appears as a top-right outlier, indicating that it captures a disproportionately high amount of energy even among other subgraphs of similar density. This strong alignment provides the empirical justification for using the energy ranking as a powerful guide in the discovery of quasi-cliques.

\subsection{Parameter Sensitivity}
\label{sec:sensitivity}

\begin{figure}[htbp]
    \centering
    \includegraphics[width=0.9\linewidth]{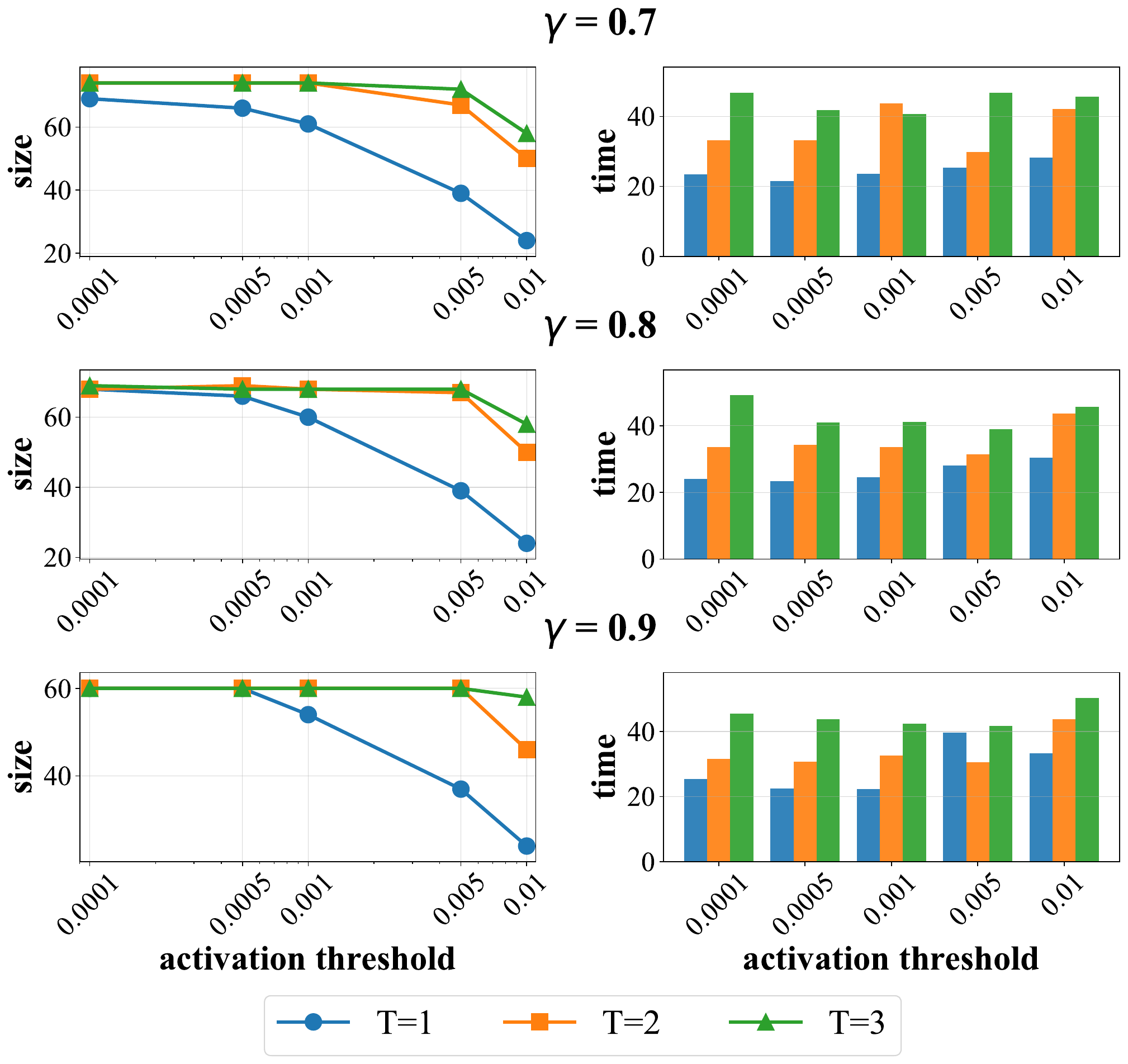}
    \caption{Sensitivity of EDQC to $(T,\theta)$ on \texttt{web-Google} under three density thresholds $\gamma$. Left: quasi-clique size; right: runtime (log-scale $\theta$).}
    \label{fig:param-large}
\end{figure}

We studied the sensitivity of EDQC to the diffusion rounds $T$ and the activation threshold $\theta$ on three datasets of different scales (\texttt{ego-facebook}, \texttt{rgg\_n\_2\_19\_s0}, and \texttt{web-Google}). We tested $T \in \{1,2,3\}$ and $\theta \in \{10^{-4}, 5 \times 10^{-4}, 10^{-3}, 5 \times 10^{-3}, 10^{-2}\}$ under three values of $\gamma$. Due to space limits, Figure~\ref{fig:param-large} reports results on \texttt{web-Google}.

Overall, EDQC is stable over a broad range of $(T,\theta)$. Increasing $T$ from $1$ to $2$ typically improves solution quality, while $T=3$ offers little additional gain but noticeably increases runtime. Smaller thresholds ($10^{-4}$ to $10^{-3}$) yield similar solution quality, whereas larger values ($\theta \ge 5 \times 10^{-3}$) tend to over-prune and slightly degrade performance, especially at higher $\gamma$. 

\section{Conclusion} \label{sec:conclusion}
This paper proposes EDQC, an effective framework for cohesive group discovery under explicit edge-density constraints. 
EDQC leverages a lightweight adaptive energy diffusion process to rank vertices and localize candidate regions in interaction graphs; guided by this ranking, it performs conductance-based extraction and density-feasible refinement to obtain a $\gamma$-quasi-clique that satisfies the target threshold. 
This design aims to alleviate practical limitations observed in existing heuristics, including sensitivity to randomness and the lack of explicit density guarantees.

Extensive experiments on 75 real-world graphs across varying density thresholds show that EDQC identifies larger mean $\gamma$-quasi-cliques in the vast majority of cases, with consistently lower variance across runs while maintaining competitive runtime, and statistical tests support the significance of these improvements. 
Overall, EDQC offers a robust primitive for cohesive group discovery in interaction graphs, aligning with graph-based recommender systems where explicitly constrained cohesive groups underpin tasks such as social recommendation, bundle discovery, and community-aware modeling. 
Future work includes extending EDQC to weighted and attributed graphs and integrating it into end-to-end recommender pipelines to quantify task-level impact.

\bibliographystyle{ACM-Reference-Format}
\bibliography{sample-base}


\begin{thebibliography}{38}


\ifx \showCODEN    \undefined \def \showCODEN     #1{\unskip}     \fi
\ifx \showISBNx    \undefined \def \showISBNx     #1{\unskip}     \fi
\ifx \showISBNxiii \undefined \def \showISBNxiii  #1{\unskip}     \fi
\ifx \showISSN     \undefined \def \showISSN      #1{\unskip}     \fi
\ifx \showLCCN     \undefined \def \showLCCN      #1{\unskip}     \fi
\ifx \shownote     \undefined \def \shownote      #1{#1}          \fi
\ifx \showarticletitle \undefined \def \showarticletitle #1{#1}   \fi
\ifx \showURL      \undefined \def \showURL       {\relax}        \fi
\providecommand\bibfield[2]{#2}
\providecommand\bibinfo[2]{#2}
\providecommand\natexlab[1]{#1}
\providecommand\showeprint[2][]{arXiv:#2}

\bibitem[Abello et~al\mbox{.}(2002)]%
        {abello2002massive}
\bibfield{author}{\bibinfo{person}{James Abello}, \bibinfo{person}{Mauricio~GC
  Resende}, {and} \bibinfo{person}{Sandra Sudarsky}.}
  \bibinfo{year}{2002}\natexlab{}.
\newblock \showarticletitle{Massive quasi-clique detection}. In
  \bibinfo{booktitle}{\emph{Latin American symposium on theoretical
  informatics}}. Springer, \bibinfo{pages}{598--612}.
\newblock


\bibitem[Alina~Christensen and Schiaffino(2014)]%
        {alina2014social}
\bibfield{author}{\bibinfo{person}{Ingrid Alina~Christensen} {and}
  \bibinfo{person}{Silvia Schiaffino}.} \bibinfo{year}{2014}\natexlab{}.
\newblock \showarticletitle{Social influence in group recommender systems}.
\newblock \bibinfo{journal}{\emph{Online Information Review}}
  \bibinfo{volume}{38}, \bibinfo{number}{4} (\bibinfo{year}{2014}),
  \bibinfo{pages}{524--542}.
\newblock


\bibitem[Andersen et~al\mbox{.}(2006)]%
        {andersen2006local}
\bibfield{author}{\bibinfo{person}{Reid Andersen}, \bibinfo{person}{Fan Chung},
  {and} \bibinfo{person}{Kevin Lang}.} \bibinfo{year}{2006}\natexlab{}.
\newblock \showarticletitle{Local graph partitioning using pagerank vectors}.
  In \bibinfo{booktitle}{\emph{2006 47th annual IEEE symposium on foundations
  of computer science (FOCS'06)}}. IEEE, \bibinfo{pages}{475--486}.
\newblock


\bibitem[Chen et~al\mbox{.}(2021)]%
        {chen2021nuqclq}
\bibfield{author}{\bibinfo{person}{Jiejiang Chen}, \bibinfo{person}{Shaowei
  Cai}, \bibinfo{person}{Shiwei Pan}, \bibinfo{person}{Yiyuan Wang},
  \bibinfo{person}{Qingwei Lin}, \bibinfo{person}{Mengyu Zhao}, {and}
  \bibinfo{person}{Minghao Yin}.} \bibinfo{year}{2021}\natexlab{}.
\newblock \showarticletitle{NuQClq: An effective local search algorithm for
  maximum quasi-clique problem}. In \bibinfo{booktitle}{\emph{Proceedings of
  the AAAI Conference on Artificial Intelligence}}, Vol.~\bibinfo{volume}{35}.
  \bibinfo{pages}{12258--12266}.
\newblock


\bibitem[Chung(2007)]%
        {chung2007heat}
\bibfield{author}{\bibinfo{person}{Fan Chung}.}
  \bibinfo{year}{2007}\natexlab{}.
\newblock \showarticletitle{The heat kernel as the pagerank of a graph}.
\newblock \bibinfo{journal}{\emph{Proceedings of the National Academy of
  Sciences}} \bibinfo{volume}{104}, \bibinfo{number}{50}
  (\bibinfo{year}{2007}), \bibinfo{pages}{19735--19740}.
\newblock


\bibitem[Dem{\v{s}}ar(2006)]%
        {demvsar2006statistical}
\bibfield{author}{\bibinfo{person}{Janez Dem{\v{s}}ar}.}
  \bibinfo{year}{2006}\natexlab{}.
\newblock \showarticletitle{Statistical comparisons of classifiers over
  multiple data sets}.
\newblock \bibinfo{journal}{\emph{Journal of Machine learning research}}
  \bibinfo{volume}{7}, \bibinfo{number}{Jan} (\bibinfo{year}{2006}),
  \bibinfo{pages}{1--30}.
\newblock


\bibitem[Djeddi et~al\mbox{.}(2019)]%
        {djeddi2019extension}
\bibfield{author}{\bibinfo{person}{Youcef Djeddi}, \bibinfo{person}{Hacene
  Ait~Haddadene}, {and} \bibinfo{person}{Nabil Belacel}.}
  \bibinfo{year}{2019}\natexlab{}.
\newblock \showarticletitle{An extension of adaptive multi-start tabu search
  for the maximum quasi-clique problem}.
\newblock \bibinfo{journal}{\emph{Computers \& Industrial Engineering}}
  \bibinfo{volume}{132} (\bibinfo{year}{2019}), \bibinfo{pages}{280--292}.
\newblock


\bibitem[Friedman(1937)]%
        {friedman1937use}
\bibfield{author}{\bibinfo{person}{Milton Friedman}.}
  \bibinfo{year}{1937}\natexlab{}.
\newblock \showarticletitle{The use of ranks to avoid the assumption of
  normality implicit in the analysis of variance}.
\newblock \bibinfo{journal}{\emph{Journal of the american statistical
  association}} \bibinfo{volume}{32}, \bibinfo{number}{200}
  (\bibinfo{year}{1937}), \bibinfo{pages}{675--701}.
\newblock


\bibitem[Gleich et~al\mbox{.}(2015)]%
        {gleich2015multilinear}
\bibfield{author}{\bibinfo{person}{David~F Gleich}, \bibinfo{person}{Lek-Heng
  Lim}, {and} \bibinfo{person}{Yongyang Yu}.} \bibinfo{year}{2015}\natexlab{}.
\newblock \showarticletitle{Multilinear pagerank}.
\newblock \bibinfo{journal}{\emph{SIAM J. Matrix Anal. Appl.}}
  \bibinfo{volume}{36}, \bibinfo{number}{4} (\bibinfo{year}{2015}),
  \bibinfo{pages}{1507--1541}.
\newblock


\bibitem[Guan et~al\mbox{.}(2021)]%
        {guan2021community}
\bibfield{author}{\bibinfo{person}{Jiewen Guan}, \bibinfo{person}{Xin Huang},
  {and} \bibinfo{person}{Bilian Chen}.} \bibinfo{year}{2021}\natexlab{}.
\newblock \showarticletitle{Community-aware social recommendation: A unified
  SCSVD framework}.
\newblock \bibinfo{journal}{\emph{IEEE Transactions on Knowledge and Data
  Engineering}} \bibinfo{volume}{35}, \bibinfo{number}{3}
  (\bibinfo{year}{2021}), \bibinfo{pages}{2379--2393}.
\newblock


\bibitem[Kloster and Gleich(2014)]%
        {kloster2014heat}
\bibfield{author}{\bibinfo{person}{Kyle Kloster} {and} \bibinfo{person}{David~F
  Gleich}.} \bibinfo{year}{2014}\natexlab{}.
\newblock \showarticletitle{Heat kernel based community detection}. In
  \bibinfo{booktitle}{\emph{Proceedings of the 20th ACM SIGKDD international
  conference on Knowledge discovery and data mining}}.
  \bibinfo{pages}{1386--1395}.
\newblock


\bibitem[Konar and Sidiropoulos(2024)]%
        {konar2024optimal}
\bibfield{author}{\bibinfo{person}{Aritra Konar} {and}
  \bibinfo{person}{Nicholas~D Sidiropoulos}.} \bibinfo{year}{2024}\natexlab{}.
\newblock \showarticletitle{Optimal quasi-clique: hardness, equivalence with
  densest-k-subgraph, and quasi-partitioned community mining}. In
  \bibinfo{booktitle}{\emph{Proceedings of the AAAI Conference on Artificial
  Intelligence}}, Vol.~\bibinfo{volume}{38}. \bibinfo{pages}{8608--8616}.
\newblock


\bibitem[Leskovec and Krevl(2014)]%
        {snapnets}
\bibfield{author}{\bibinfo{person}{Jure Leskovec} {and} \bibinfo{person}{Andrej
  Krevl}.} \bibinfo{year}{2014}\natexlab{}.
\newblock \bibinfo{title}{{SNAP Datasets}: {Stanford} Large Network Dataset
  Collection}.
\newblock


\bibitem[Lin et~al\mbox{.}(2017)]%
        {lin2017reduction}
\bibfield{author}{\bibinfo{person}{Jinkun Lin}, \bibinfo{person}{Shaowei Cai},
  \bibinfo{person}{Chuan Luo}, {and} \bibinfo{person}{Kaile Su}.}
  \bibinfo{year}{2017}\natexlab{}.
\newblock \showarticletitle{A Reduction based Method for Coloring Very Large
  Graphs}. In \bibinfo{booktitle}{\emph{Proceedings of the 26th International
  Joint Conference on Artificial Intelligence}}. \bibinfo{pages}{517--523}.
\newblock


\bibitem[Liu and Wong(2008)]%
        {liu2008effective}
\bibfield{author}{\bibinfo{person}{Guimei Liu} {and} \bibinfo{person}{Limsoon
  Wong}.} \bibinfo{year}{2008}\natexlab{}.
\newblock \showarticletitle{Effective pruning techniques for mining
  quasi-cliques}. In \bibinfo{booktitle}{\emph{Joint European conference on
  machine learning and knowledge discovery in databases}}. Springer,
  \bibinfo{pages}{33--49}.
\newblock


\bibitem[Liu et~al\mbox{.}(2024)]%
        {liu2024optimization}
\bibfield{author}{\bibinfo{person}{Shuhong Liu}, \bibinfo{person}{Jincheng
  Zhou}, \bibinfo{person}{Dan Wang}, \bibinfo{person}{Zaijun Zhang}, {and}
  \bibinfo{person}{Mingjie Lei}.} \bibinfo{year}{2024}\natexlab{}.
\newblock \showarticletitle{An optimization algorithm for maximum quasi-clique
  problem based on information feedback model}.
\newblock \bibinfo{journal}{\emph{PeerJ Computer Science}}
  \bibinfo{volume}{10} (\bibinfo{year}{2024}), \bibinfo{pages}{e2173}.
\newblock


\bibitem[Mokken et~al\mbox{.}(1979)]%
        {mokken1979cliques}
\bibfield{author}{\bibinfo{person}{Robert~J Mokken} {et~al\mbox{.}}}
  \bibinfo{year}{1979}\natexlab{}.
\newblock \showarticletitle{Cliques, clubs and clans}.
\newblock \bibinfo{journal}{\emph{Quality \& Quantity}} \bibinfo{volume}{13},
  \bibinfo{number}{2} (\bibinfo{year}{1979}), \bibinfo{pages}{161--173}.
\newblock


\bibitem[Nguyen et~al\mbox{.}(2024)]%
        {nguyen2024bundle}
\bibfield{author}{\bibinfo{person}{Huy-Son Nguyen}, \bibinfo{person}{Tuan-Nghia
  Bui}, \bibinfo{person}{Long-Hai Nguyen}, \bibinfo{person}{Hung Hoang},
  \bibinfo{person}{Cam-Van Thi~Nguyen}, \bibinfo{person}{Hoang-Quynh Le}, {and}
  \bibinfo{person}{Duc-Trong Le}.} \bibinfo{year}{2024}\natexlab{}.
\newblock \showarticletitle{Bundle Recommendation with Item-Level
  Causation-Enhanced Multi-view Learning}. In \bibinfo{booktitle}{\emph{Joint
  European Conference on Machine Learning and Knowledge Discovery in
  Databases}}. Springer, \bibinfo{pages}{324--341}.
\newblock


\bibitem[Oliveira et~al\mbox{.}(2013)]%
        {oliveira2013construction}
\bibfield{author}{\bibinfo{person}{AB Oliveira}, \bibinfo{person}{A Plastino},
  {and} \bibinfo{person}{CC Ribeiro}.} \bibinfo{year}{2013}\natexlab{}.
\newblock \showarticletitle{Construction heuristics for the maximum cardinality
  quasi-clique problem}. In \bibinfo{booktitle}{\emph{10th Metaheuristics
  International Conference. Singapore}}, Vol.~\bibinfo{volume}{84}.
\newblock


\bibitem[Page et~al\mbox{.}(1999)]%
        {page1999pagerank}
\bibfield{author}{\bibinfo{person}{Lawrence Page}, \bibinfo{person}{Sergey
  Brin}, \bibinfo{person}{Rajeev Motwani}, {and} \bibinfo{person}{Terry
  Winograd}.} \bibinfo{year}{1999}\natexlab{}.
\newblock \bibinfo{booktitle}{\emph{The PageRank citation ranking: Bringing
  order to the web.}}
\newblock \bibinfo{type}{{T}echnical {R}eport}. \bibinfo{institution}{Stanford
  infolab}.
\newblock


\bibitem[Pang et~al\mbox{.}(2024)]%
        {pang2024similarity}
\bibfield{author}{\bibinfo{person}{Jiayang Pang}, \bibinfo{person}{Chenhao Ma},
  {and} \bibinfo{person}{Yixiang Fang}.} \bibinfo{year}{2024}\natexlab{}.
\newblock \showarticletitle{A Similarity-based Approach for Efficient Large
  Quasi-clique Detection}. In \bibinfo{booktitle}{\emph{Proceedings of the ACM
  Web Conference 2024}}. \bibinfo{pages}{401--409}.
\newblock


\bibitem[Pastukhov et~al\mbox{.}(2018)]%
        {pastukhov2018maximum}
\bibfield{author}{\bibinfo{person}{Grigory Pastukhov},
  \bibinfo{person}{Alexander Veremyev}, \bibinfo{person}{Vladimir Boginski},
  {and} \bibinfo{person}{Oleg~A Prokopyev}.} \bibinfo{year}{2018}\natexlab{}.
\newblock \showarticletitle{On maximum degree-based-quasi-clique problem:
  Complexity and exact approaches}.
\newblock \bibinfo{journal}{\emph{Networks}} \bibinfo{volume}{71},
  \bibinfo{number}{2} (\bibinfo{year}{2018}), \bibinfo{pages}{136--152}.
\newblock


\bibitem[Pattillo et~al\mbox{.}(2013)]%
        {pattillo2013maximum}
\bibfield{author}{\bibinfo{person}{Jeffrey Pattillo},
  \bibinfo{person}{Alexander Veremyev}, \bibinfo{person}{Sergiy Butenko}, {and}
  \bibinfo{person}{Vladimir Boginski}.} \bibinfo{year}{2013}\natexlab{}.
\newblock \showarticletitle{On the maximum quasi-clique problem}.
\newblock \bibinfo{journal}{\emph{Discrete Applied Mathematics}}
  \bibinfo{volume}{161}, \bibinfo{number}{1-2} (\bibinfo{year}{2013}),
  \bibinfo{pages}{244--257}.
\newblock


\bibitem[Pei et~al\mbox{.}(2005)]%
        {pei2005mining}
\bibfield{author}{\bibinfo{person}{Jian Pei}, \bibinfo{person}{Daxin Jiang},
  {and} \bibinfo{person}{Aidong Zhang}.} \bibinfo{year}{2005}\natexlab{}.
\newblock \showarticletitle{On mining cross-graph quasi-cliques}. In
  \bibinfo{booktitle}{\emph{Proceedings of the eleventh ACM SIGKDD
  international conference on Knowledge discovery in data mining}}.
  \bibinfo{pages}{228--238}.
\newblock


\bibitem[Rahman et~al\mbox{.}(2024)]%
        {rahman2024fast}
\bibfield{author}{\bibinfo{person}{Ahsanur Rahman}, \bibinfo{person}{Kalyan
  Roy}, \bibinfo{person}{Ramiza Maliha}, {and} \bibinfo{person}{Townim~Faisal
  Chowdhury}.} \bibinfo{year}{2024}\natexlab{}.
\newblock \showarticletitle{A Fast Exact Algorithm to Enumerate Maximal
  Pseudo-cliques in Large Sparse Graphs}. In
  \bibinfo{booktitle}{\emph{Proceedings of the 30th ACM SIGKDD Conference on
  Knowledge Discovery and Data Mining}}. \bibinfo{pages}{2479--2490}.
\newblock


\bibitem[Ribeiro and Riveaux(2019)]%
        {ribeiro2019exact}
\bibfield{author}{\bibinfo{person}{Celso~C Ribeiro} {and}
  \bibinfo{person}{Jos{\'e}~A Riveaux}.} \bibinfo{year}{2019}\natexlab{}.
\newblock \showarticletitle{An exact algorithm for the maximum quasi-clique
  problem}.
\newblock \bibinfo{journal}{\emph{International Transactions in Operational
  Research}} \bibinfo{volume}{26}, \bibinfo{number}{6} (\bibinfo{year}{2019}),
  \bibinfo{pages}{2199--2229}.
\newblock


\bibitem[Rossi and Ahmed(2015)]%
        {nr}
\bibfield{author}{\bibinfo{person}{Ryan~A. Rossi} {and}
  \bibinfo{person}{Nesreen~K. Ahmed}.} \bibinfo{year}{2015}\natexlab{}.
\newblock \showarticletitle{The Network Data Repository with Interactive Graph
  Analytics and Visualization}. In \bibinfo{booktitle}{\emph{AAAI}}.
\newblock


\bibitem[Rossi et~al\mbox{.}(2014)]%
        {rossi2014fast}
\bibfield{author}{\bibinfo{person}{Ryan~A Rossi}, \bibinfo{person}{David~F
  Gleich}, \bibinfo{person}{Assefaw~H Gebremedhin}, {and}
  \bibinfo{person}{Md~Mostofa~Ali Patwary}.} \bibinfo{year}{2014}\natexlab{}.
\newblock \showarticletitle{Fast maximum clique algorithms for large graphs}.
  In \bibinfo{booktitle}{\emph{Proceedings of the 23rd International Conference
  on World Wide Web}}. \bibinfo{pages}{365--366}.
\newblock


\bibitem[Saito et~al\mbox{.}(2007)]%
        {saito2007large}
\bibfield{author}{\bibinfo{person}{Hiroo Saito}, \bibinfo{person}{Masashi
  Toyoda}, \bibinfo{person}{Masaru Kitsuregawa}, {and}
  \bibinfo{person}{Kazuyuki Aihara}.} \bibinfo{year}{2007}\natexlab{}.
\newblock \showarticletitle{A large-scale study of link spam detection by graph
  algorithms}. In \bibinfo{booktitle}{\emph{Proceedings of the 3rd
  international workshop on Adversarial information retrieval on the web}}.
  \bibinfo{pages}{45--48}.
\newblock


\bibitem[Schafer et~al\mbox{.}(2001)]%
        {schafer2001commerce}
\bibfield{author}{\bibinfo{person}{J~Ben Schafer}, \bibinfo{person}{Joseph~A
  Konstan}, {and} \bibinfo{person}{John Riedl}.}
  \bibinfo{year}{2001}\natexlab{}.
\newblock \showarticletitle{E-commerce recommendation applications}.
\newblock \bibinfo{journal}{\emph{Data mining and knowledge discovery}}
  \bibinfo{volume}{5}, \bibinfo{number}{1} (\bibinfo{year}{2001}),
  \bibinfo{pages}{115--153}.
\newblock


\bibitem[Seidman and Foster(1978)]%
        {seidman1978graph}
\bibfield{author}{\bibinfo{person}{Stephen~B Seidman} {and}
  \bibinfo{person}{Brian~L Foster}.} \bibinfo{year}{1978}\natexlab{}.
\newblock \showarticletitle{A graph-theoretic generalization of the clique
  concept}.
\newblock \bibinfo{journal}{\emph{Journal of Mathematical sociology}}
  \bibinfo{volume}{6}, \bibinfo{number}{1} (\bibinfo{year}{1978}),
  \bibinfo{pages}{139--154}.
\newblock


\bibitem[Sun et~al\mbox{.}(2025)]%
        {sunnumds}
\bibfield{author}{\bibinfo{person}{Rui Sun}, \bibinfo{person}{Zhaohui Liu},
  \bibinfo{person}{Yiyuan Wang}, \bibinfo{person}{Han Xiao},
  \bibinfo{person}{Jiangnan Li}, {and} \bibinfo{person}{Jiejiang Chen}.}
  \bibinfo{year}{2025}\natexlab{}.
\newblock \showarticletitle{NuMDS: An Efficient Local Search Algorithm for
  Minimum Dominating Set Problem}. In \bibinfo{booktitle}{\emph{Proceedings of
  the 34th International Joint Conference on Artificial Intelligence}}.
\newblock


\bibitem[Tsourakakis et~al\mbox{.}(2013)]%
        {tsourakakis2013denser}
\bibfield{author}{\bibinfo{person}{Charalampos Tsourakakis},
  \bibinfo{person}{Francesco Bonchi}, \bibinfo{person}{Aristides Gionis},
  \bibinfo{person}{Francesco Gullo}, {and} \bibinfo{person}{Maria Tsiarli}.}
  \bibinfo{year}{2013}\natexlab{}.
\newblock \showarticletitle{Denser than the densest subgraph: extracting
  optimal quasi-cliques with quality guarantees}. In
  \bibinfo{booktitle}{\emph{Proceedings of the 19th ACM SIGKDD international
  conference on Knowledge discovery and data mining}}.
  \bibinfo{pages}{104--112}.
\newblock


\bibitem[Uno(2010)]%
        {uno2010efficient}
\bibfield{author}{\bibinfo{person}{Takeaki Uno}.}
  \bibinfo{year}{2010}\natexlab{}.
\newblock \showarticletitle{An efficient algorithm for solving pseudo clique
  enumeration problem}.
\newblock \bibinfo{journal}{\emph{Algorithmica}}  \bibinfo{volume}{56}
  (\bibinfo{year}{2010}), \bibinfo{pages}{3--16}.
\newblock


\bibitem[Xia et~al\mbox{.}(2025)]%
        {xia2025maximum}
\bibfield{author}{\bibinfo{person}{Hongbo Xia}, \bibinfo{person}{Kaiqiang Yu},
  \bibinfo{person}{Shengxin Liu}, \bibinfo{person}{Cheng Long}, {and}
  \bibinfo{person}{Xun Zhou}.} \bibinfo{year}{2025}\natexlab{}.
\newblock \showarticletitle{Maximum Degree-Based Quasi-Clique Search via an
  Iterative Framework}.
\newblock \bibinfo{journal}{\emph{Proceedings of the 31th ACM SIGKDD Conference
  on Knowledge Discovery and Data Mining}} (\bibinfo{year}{2025}).
\newblock


\bibitem[Zheng et~al\mbox{.}(2023)]%
        {zheng2023farsighted}
\bibfield{author}{\bibinfo{person}{Jiongzhi Zheng}, \bibinfo{person}{Kun He},
  {and} \bibinfo{person}{Jianrong Zhou}.} \bibinfo{year}{2023}\natexlab{}.
\newblock \showarticletitle{Farsighted probabilistic sampling: A general
  strategy for boosting local search MaxSAT solvers}. In
  \bibinfo{booktitle}{\emph{Proceedings of the AAAI Conference on Artificial
  Intelligence}}, Vol.~\bibinfo{volume}{37}. \bibinfo{pages}{4132--4139}.
\newblock


\bibitem[Zhou et~al\mbox{.}(2020)]%
        {zhou2020opposition}
\bibfield{author}{\bibinfo{person}{Qing Zhou}, \bibinfo{person}{Una Benlic},
  {and} \bibinfo{person}{Qinghua Wu}.} \bibinfo{year}{2020}\natexlab{}.
\newblock \showarticletitle{An opposition-based memetic algorithm for the
  maximum quasi-clique problem}.
\newblock \bibinfo{journal}{\emph{European Journal of Operational Research}}
  \bibinfo{volume}{286}, \bibinfo{number}{1} (\bibinfo{year}{2020}),
  \bibinfo{pages}{63--83}.
\newblock


\bibitem[Zhou(2025)]%
        {zhou2025robust}
\bibfield{author}{\bibinfo{person}{Xinyi Zhou}.}
  \bibinfo{year}{2025}\natexlab{}.
\newblock \showarticletitle{Robust Personalized Movie Recommendation System
  Based on Embedding Technology}. In \bibinfo{booktitle}{\emph{Proceedings of
  the 2025 3rd International Conference on Internet of Things and Cloud
  Computing Technology}}. \bibinfo{pages}{354--358}.
\newblock


\end{thebibliography}

\end{document}